\documentclass[reprint,aps,prx,amsmath,amssymb,longbibliography, notitlepage, footinbib,superscriptaddress]{revtex4-1}
\usepackage{amsmath, amssymb, bm, graphicx}
\usepackage{graphics}
\usepackage{float} 
\usepackage{mathtools}
\usepackage[stable]{footmisc}
\usepackage{soul}
\usepackage[colorlinks, linkcolor= blue, citecolor = blue, urlcolor=blue]{hyperref}
\usepackage{pifont}
\usepackage{algorithm}
\usepackage{chemformula}

\newcommand{\cmark}{\ding{51}}%
\newcommand{\xmark}{\ding{55}}%
\usepackage{physics}
\def\be{\begin{equation}}
\def\ee{\end{equation}}
\def \bea{\begin{eqnarray}}
\def \eea{\end{eqnarray}}
\def \nn{\nonumber}

\begin{document}

\title{Odd-parity longitudinal magnetoconductivity in time-reversal symmetry broken materials}

\author{Sunit Das}
\thanks{Sunit and Akash contributed equally to this work.}
\affiliation{Department of Physics, Indian Institute of Technology Kanpur, Kanpur 208016, India}
\author{Akash Adhikary}
\thanks{Sunit and Akash contributed equally to this work.}
\affiliation{Department of Physics, Indian Institute of Technology Kanpur, Kanpur 208016, India}
\author{Divya Sahani} 
\affiliation{Department of Physics, Indian Institute of Science, Bangalore 560012, India}
\author{Aveek Bid}
\email{aveek@iisc.ac.in}
\affiliation{Department of Physics, Indian Institute of Science, Bangalore 560012, India}
\author{Amit Agarwal}
\email{amitag@iitk.ac.in}
\affiliation{Department of Physics, Indian Institute of Technology Kanpur, Kanpur 208016, India}

\begin{abstract}
Magnetotransport measurements are a sensitive probe of symmetry and electronic structure in quantum materials. While conventional metals exhibit longitudinal magnetoconductivity that is even in a magnetic field ($B$) for small $B$, we show that magnetic materials which intrinsically break time-reversal symmetry (TRS) show an \emph{odd-parity magnetoconductivity} (OMC), with a leading linear-$B$ response. Using semiclassical transport theory, we derive explicit expressions for the longitudinal and transverse conductivities and identify their origin in Berry curvature and orbital magnetic moment. Crystalline symmetry analysis shows that longitudinal OMC follows the same point-group constraints as the anomalous Hall effect, while transverse OMC obeys distinct rules, providing an independent probe of TRS breaking. In the large $B$ quantum oscillation regime, we uncover both odd- and even-$B$ contributions, demonstrating OMC beyond the semiclassical picture. Explicit calculations in valley-polarized gapped graphene show that OMC peaks near the band edges, vanish in the band gap and follow the temperature dependence of the magnetic order parameter. Our results explain the odd-parity magnetoresistance recently observed in magnetized graphene and establish OMC as a robust transport signature of intrinsic TRS breaking in metals.
\end{abstract}

\maketitle

\begin{figure}[t]
    \centering
    \includegraphics[width=0.8\linewidth]{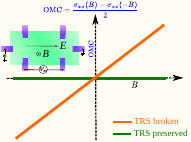}
    \caption{Schematic of longitudinal odd-parity magnetoconductivity (OMC) in the low-field limit. In time-reversal symmetric (TRS) materials, Onsager reciprocity forbids OMC, leaving only even-$B$ responses. In contrast, intrinsic TRS-breaking in magnetic systems produces a leading $B$-linear longitudinal response (OMC). Thus, a finite OMC serves as a direct transport signature of intrinsic TRS breaking.}
    \label{fig1}
\end{figure}

% \clearpage
\section{Introduction}
Magnetotransport measurements are a powerful probe of the symmetry, topology, and Fermi surface geometry of quantum materials~\cite{Lu2017}. In conventional metals, semiclassical magnetoresistance varies quadratically with magnetic field ($B$) for small $B$ and provides information about carrier mobility and charge-compensation~\cite{AshcroftMermin, Pippard}. In the high-field regime, quantum oscillations in resistance reveal effective masses, Fermi surface area, and topological phases~\cite{lifshitz1956theory, lifshitz1958theory, Novoselov_nat05, Zhang_nat05,glazman_fermiology23, sharlai_22wsm, gao_PNAS17, Fuch_scipost18, das2024nonlinearlandaufandiagram, Kumari_afm25, Adak_nrm24}. More recently, band-geometric quantities such as Berry curvature and orbital magnetic moment have been shown to drive unconventional magnetotransport phenomena, including the chiral anomaly-induced negative magnetoresistance~\cite{Spivak_prb13, Spivak_prb16, Chen_prx15,Zhang_nc16, Kamal_prb20, Kamal_prb20chiral, Sunit_prb23, mandal_prb22}, anisotropic magnetoresistance~\cite{mcguire2003anisotropic, Wang_prb05, Nandy_prl17, Fina_nc14, Chandra_prb18, Ghorai_prl25, Rahul_prb25}. Notably, these responses are typically even in the applied magnetic field.

The even-$B$ nature of longitudinal and odd-$B$ nature of transverse responses follow directly from Onsager’s reciprocity relation, $\sigma_{ab}(B)=\sigma_{ba}(-B)$~\cite{Onsager_31, Mazur_53}. This enforces longitudinal conductivities to be even in $B$, while the antisymmetric off-diagonal elements ($a \leftrightarrow b$ with $a \neq b$) describe the conventional Lorentz Hall effect. However, in systems with intrinsically broken time-reversal symmetry (TRS), Onsager’s relation generalizes to $\sigma_{ab}(B, M)=\sigma_{ba}(-B,-M)$, with $M$ being the intrinsic TRS-breaking parameter such as magnetization~\cite{Buttiker_prb12}. As a result, longitudinal conductivities can acquire odd-$B$ contributions in magnetic materials, which intrinsically break TRS. 

Odd-$B$ magnetoconductivity has indeed been observed in several TRS-broken systems and studied theoretically~\cite{Sahani_prl24, Cortijo_prb16, Tiwari_prb17, Das_prb19_electrical, Kamal_prb19_thermo, tobius_prb20, Nui_prb20, Zyuzin_prb21, Wu_prl21, Orenstein_23linear, albarakati2019antisymmetric, Wang_NC20, niu2021antisymmetric, fujita2015odd,jiang2021chirality,takiguchi2022giant,moubah2014antisymmetric, Tanaka_25}. However, its microscopic origin and crystalline symmetry restrictions remain underexplored. More importantly, a comprehensive study of the odd-$B$ longitudinal magnetoconductivity across both the low-field semiclassical and high-field quantum oscillation regimes is still lacking. 

Motivated by these, in this paper, we systematically study \emph{odd-parity magnetoconductivity} (OMC). Using semiclassical transport formalism, we derive explicit expressions for the $B$-linear longitudinal and transverse responses, tracing their origin to Berry curvature (BC) and orbital magnetic moment (OMM). Our symmetry analysis reveals that the longitudinal OMC and anomalous Hall conductivity (AHC) share the same point-group symmetry restrictions and should generally appear together. However, the spatially symmetric transverse OMC follows distinct constraints, and it offers an independent probe of TRS breaking. Explicit calculations for valley-polarized gapped graphene demonstrate that OMC peaks near band edges and vanish inside the bandgap. Additionally, by incorporating a temperature-dependent order parameter for TRS-breaking, we show that OMC tracks the underlying magnetic order and vanishes at the critical temperature. Thus, OMC is an effective order parameter for probing magnetic phase transitions in transport experiments. 

We further extend our analysis to the quantum oscillation regime, showing that odd-parity responses persist in high fields alongside the conventional even-$B$ contributions. In the ultra-quantum limit, the OMC reduces to a universal $B$-linear behavior consistent with Abrikosov’s theory of linear magnetoresistance in systems with linearly dispersing bands~\cite{Abrikosov_prb98, Abrikosov2000}. These explain the recently observed odd-parity magnetoresistivity (OMR) in magnetized graphene~\cite {Sahani_prl24}. Our results establish OMC and its resistive counterpart, OMR, as robust transport signatures of intrinsic TRS breaking, complementary to anomalous Hall transport and directly relevant for identifying topological magnetic phases.

The rest of this paper is organized as follows. Section~\ref{Sec_2} develops expressions for the $B$-linear magnetoconductivities within semiclassical transport theory. Section~\ref{Sec_3}, presents a detailed crystalline symmetry analysis of OMC and its relation to the anomalous Hall effect. In Sec.~\ref{Sec_4}, we discuss both $B$-linear and $B$-quadratic longitudinal and transverse conductivities, highlighting their distinct scaling with scattering time. Section~\ref{Sec_5} extends the discussion to the quantum oscillation regime, where Landau quantization generates both even-$B$ and odd-$B$ responses. We conclude in Sec.~\ref{Sec_6} with a summary of the main findings and their implications.
%We summarize our findings and their implications in Sec.~\ref{Sec_6}.

% \textcolor{cyan}{Although the anomalous Hall effect(AHE) is commonly used as a definite probe of time-reversal symmetry breaking, magnetization by itself does not guarantee a nonzero AHE. Symmetry and the Berry-curvature distribution of the bands determine the intrinsic Hall response, and in some materials the AHE can be tuned to zero or be symmetry-forbidden despite finite magnetic order\cite{manna2018colossal}. OMC can be used as a hallmark of TRS breaking in those cases.}
%\textcolor{red}{We have to prove OMC can survive in those cases.}

\section{Semiclassical theory of OMC}\label{Sec_2}
We start with the derivation of $B$-linear magnetoconductivity in the presence of a weak magnetic field. We use the semiclassical equations of motion with BC and OMM corrections to show how odd-$B$ longitudinal magnetoconductivity arises. We consider a two-dimensional system with an electric field applied along the $\hat{\bm x}$-direction (${\bm E}=E\hat{\bm x}$) and the magnetic field applied along $\hat{\bm z}$-direction (${\bm B}=B\hat{\bm z}$). The equations of motion for electronic wave packets, including the BC ($\bm \Omega$) and OMM ($\bm m$) under weak electric and magnetic fields for the $n$th band, are given by~\cite{Niu_rmp10}
\begin{subequations}
\bea 
\dot{\bm r}_n & =&  {\cal D}_n \left[\tilde{\bm v}_n +\frac{e}{\hbar} ({\bm E} \times {\bm \Omega}_n) \right] , \label{eom_r}\\
\hbar\dot{\bm k }_n & = & {\cal D}_n \left[-e{\bm E} - e{ (\tilde{\bm v}_n} \times {\bm B}) \right]. \label{eom_k}
\eea
\end{subequations}
Here, `$-e$' is the electronic charge. In the above equations, ${\cal D}_n \equiv 1/(1 + \frac{e}{\hbar}{\bm \Omega}_n  \cdot {\bm B})$ is the phase-space factor, which modifies the invariant phase-space volume according to $[d{\bm k}] \to [d{\bm k}] {\cal D}_n^{-1}$~\cite{xiao_prl05}. The modified band energy in the presence of OMM~\cite{Burgos_AP24}, $\tilde{\varepsilon}_n=\varepsilon_n-{\bm m}_n \cdot{\bm B}$ gives rise to a correction to the band velocity, $\tilde{\bm v}_n =(1/\hbar) {\bm \nabla}_{\bm k} \tilde{\varepsilon}_n={\bm v}^0_n + {\bm v}^m_n $, where ${\bm v}_n^0 = (1/\hbar) {\bm \nabla}_{\bm k} {\varepsilon}_n $ and ${\bm v}^m_n = (1/\hbar) {\bm \nabla}_{\bm k} (-{\bm m}_n \cdot{\bm B})$.

Semiclassically, the current density is expressed as ${\bm j} = -e \sum_n \int [d {\bm k}]{\cal D}^{-1}_n {\dot{\bm r}}_n g_{n\bm k}$. Here, $[d{\bm k}]\equiv \frac{d^2 k}{(2\pi)^2}$, and $g_{n\bm k}$ is the non-equilibrium distribution function calculated from the Boltzmann transport equation. The linear response conductivity $\sigma_{ab}$ is evaluated as $j_a=\sigma_{ab}E_b$. In the non-equilibrium steady-state, using the relaxation time approximation, the $g_{n\bm k}$ is obtained to be
\bea \label{distr_fn}
g_{n{\bm k}} &=& f_{n}^0 +
\sum_{l=0}^{\infty} \left( \frac{e\tau}{\hbar} {\cal D}_n(\bm{\tilde v}_n \times \bm{B})\cdot\bm\nabla_{\bm{k}} \right)^l  \nn \\
&& \times \left[e \tau {\cal D}_n ( \bm{\tilde v}_n \cdot \bm{E} ) \frac{\partial f_n^0}{\partial \varepsilon_n} \right].
\eea
Here, $f_n^0=1/[1+e^{\beta_T({\tilde \varepsilon}_n-\mu)}]$ is the equilibrium Fermi-Dirac distribution function with the inverse temperature $\beta_T =1/k_BT$. $\tau$ is the relaxation time, which we assume to be constant. Using Eq.~\eqref{distr_fn}, we obtain the $\sigma_{ab}$ to linear order in magnetic field strength as follows 
\begin{widetext}
\begin{subequations}
\bea \label{sigma_B}
\sigma_{ab}(B) & = & \sigma_{ab}^{\rm O}(B) + \sigma_{ab}^{\rm L}(B) + \sigma_{ab}^{\rm OMC}(B) ~~\text{with} \label{sigma_B_tot} \\ 
\sigma_{ab}^{\rm O}(B) &=& \frac{e^2}{\hbar} \epsilon_{abc} \int_{n,{\bm k}}  {\Omega}^c ({\bm m}\cdot {\bm B}) (\partial_{\varepsilon} f_n^0)~, \quad~ \sigma_{ab}^{\rm L}(B) = - \frac{e^3 \tau^2}{\hbar} \int_{n,{\bm k}}  { v}_{a}^0 ({\bm v}^0 \times {\bm B}) \cdot {\bm \nabla}_{\bm k}\left[{ v}_{b}^0 ~ (\partial_{\varepsilon} f_n^0) 
 \right],  \label{sigma_B_hall} \\
 \sigma_{ab}^{\rm OMC}(B) &=& - e^2 \tau \int_{n,{\bm k}} \left[ ({v}_{a}^m {v}_{b}^0 + { v}_{a}^0  v_{b}^m )  \partial_{\varepsilon} f_n^0 -  { v}_{a}^0 {v}_{b}^0 \partial_{\varepsilon}\left( {\bm m}\cdot {\bm B}~ \partial_{\varepsilon} f_n^0 \right
)\right] + \frac{e^3\tau}{\hbar} \int_{n,{\bm k}} ({\bm \Omega}\cdot {\bm B}) {v}_{a}^0 { v}_{b}^0 (\partial_{\varepsilon} f_n^0)~. \label{sigma_B_omc}
\eea
\end{subequations}
\end{widetext}
\begin{table*}[t]
    \centering
    \caption{The crystalline point group symmetry restrictions for anomalous Hall conductivity (AHC) and the longitudinal and transverse components of the OMC. The tick-mark (\cmark) and cross-mark (\xmark) represent that the corresponding elements are symmetry allowed and forbidden, respectively. The AHC and OMC are allowed under $\cal P$ symmetry and forbidden if the system possesses $\cal T$ or $\cal PT$ symmetry. Here, ${\cal M}_{a}$, ${\cal C}_{na}$ represent  mirror, $n$-fold rotation symmetry operation along the $a$-direction for $a=\{x,y,z\}$, respectively.   }
    \renewcommand{\arraystretch}{1}
    \setlength\tabcolsep{0.24 cm}
    \begin{tabular}{c | c c c c c c c c c c c c c c c}
       \hline \hline
        Response & ${\cal M}_x$ & ${\cal M}_y$ & ${\cal M}_z$  & ${\cal C}_{2x}$ & ${\cal C}_{2y}$ & ${\cal C}_{2z}$ &  ${\cal C}_{3z}$ & ${\cal C}_{4z}$  & ${\cal M}_x \cal T$ & ${\cal M}_y \cal T$ & ${\cal M}_z {\cal T}$ & ${\cal C}_{2x}\cal T$ & ${\cal C}_{2y}\cal T$ & ${\cal C}_{2z} \cal T$  %& ${\cal C}_{4z}\cal T$ 
         \\ 
        \hline 
         $\chi_{xx;z}^{\rm OMC}$ \& $\sigma_{yx}^{\rm AH}$ &  \xmark & \xmark & \cmark & \xmark & \xmark & \cmark & \cmark & \cmark &\cmark & \cmark & \xmark & \cmark & \cmark & \xmark  \\
        % $\chi_{xx;z}^{\rm LMC}$ & \xmark & \xmark & \cmark & \xmark & \xmark & \cmark & \cmark & \cmark & \cmark & \xmark & \cmark & \cmark & \xmark & \xmark  \\
        $\chi_{yx;z}^{\rm OMC}$ & \cmark & \cmark & \cmark & \cmark & \cmark & \cmark & \xmark & \xmark & \xmark & \xmark & \xmark & \xmark & \xmark & \xmark   \\
        \hline \hline
    \end{tabular}
    \label{table_symmetry}
\end{table*}
Here, we used the shorthand $\int_{n\bm{k}} \equiv \sum_n \int [d\bm{k}]$, and omit explicit reference to the band index and momentum in the relevant quantities. The response $\sigma^{\rm O}_{ab}$ denotes the intrinsic Hall response induced by the BC and OMM~\cite{kamal_prb21}, while $\sigma^{\rm L}_{ab}$ represents the Lorentz Hall response. Both contributions are antisymmetric under the interchange of spatial indices and therefore correspond to a genuine Hall response. In contrast, the odd-parity magnetoconductivity $\sigma^{\rm OMC}_{ab}$ is symmetric under $a\leftrightarrow b$ and can contribute to both longitudinal and transverse currents. For a detailed discussion on the symmetric odd-$B$ transverse magnetoconductivity, we refer to Ref.~\cite{Orenstein_23linear}. {The odd-parity conductivity $\sigma^{\rm OMC}_{ab}$ receives three distinct microscopic contributions. First, the orbital magnetic moment shifts the band energy, thereby modifying the group velocity and producing the `velocity correction' term. Second, the OMM-modified energy shift alters the equilibrium occupation on the Fermi surface, giving an additional contribution through the distribution function. Third, the Berry curvature alters the phase-space volume in the presence of a magnetic field, resulting in the geometric `phase-space' term. These three elements together constitute the full microscopic structure of the OMC response.}

% Importantly, $\sigma^{\rm OMC}_{ab}$ is finite only in systems which are intrinsically magnetic and break TRS, even when $B=0$. In TRS-preserving systems, the $\bm{\Omega}$, $\bm{m}$, and $\bm{v}^0$ are all odd and $\bm{v}^m$ is even in momentum $\bm k$, respectively. This makes the integrals in Eq.~\eqref{sigma_B_omc} vanish. Hence, $B$-linear longitudinal magnetoconductivity can only occur in a TRS-broken state, consistent with Onsager’s reciprocity relation~\cite{Onsager_31}. 

Apart from the odd-$B$ contribution, we also have field-independent Drude conductivity and the $B^2$-dependent magnetoconductivity $\sigma_{ab}(B^2)$. Their explicit expressions are given in Appendix~\ref{app_A}. Interestingly, the $\sigma_{ab}(B^2)$ acquires multiple contributions with distinct $\tau$-scalings: $\tau^0$, $\tau$, $\tau^2$, and $\tau^3$. The $\tau^0$ and $\tau^2$ dependent terms are antisymmetric under the exchange of spatial indices {\it i.e.,} $\sigma_{ab}(B^2)=-\sigma_{ba}(B^2)$. Consequently, they only contribute to the transverse response, whereas others can support both the longitudinal and transverse response. 

Since $\sigma^{\rm OMC}_{ab}$ is finite only in TRS–broken materials, it serves as a direct signature of intrinsic TRS breaking. In experiments, however, the longitudinal magnetoresponse always contains a $B$-independent Drude term together with conventional $B^2$ contributions. The odd-parity magnetoconductivity compared to the conventional even-parity component can be quantified by the magnitude of the following ratio
\be \label{OMC_per}
{\rm OMC} \%  = \frac{\sigma^{\rm tot }_{ab} - \sigma^{\rm even}_{ab}}{\sigma^{\rm even}_{ab}} \times 100,
\ee
where $\sigma^{\rm tot}_{ab}$ is the measured total conductivity and $\sigma^{\rm even}_{ab}$ is the even-$B$ part, including the zero-field Drude contribution. This experimentally accessible ratio removes the dominant Drude and even-$B$ backgrounds, isolating the odd-parity response. 
More importantly, OMC\% can act as a dimensionless effective order parameter for the TRS broken state in transport experiments, simultaneously quantifying the odd-parity response. Its variation with system parameters such as carrier density, strain, and temperature can be used to extract valuable information about the underlying magnetic order in the system~\cite{Sahani_prl24}.

{It is also important to understand how OMC\% scales with the scattering lifetime. The odd-parity longitudinal magnetoconductivity (OMC) scales linearly with the relaxation time $\tau$, whereas the Drude contribution and the even-$B$ longitudinal conductivity scale as $\tau$ and $\tau^{3}$, respectively (see Appendix~\ref{app_A}). The relative measure ${\rm OMC}\%$ therefore scales approximately as $\tau^{-2}$. Thus, while ultra-clean samples exhibit a larger absolute OMC signal, their relative OMC\% is reduced, whereas samples with moderate disorder (intermediate $\tau$) provide more favourable conditions for observing a pronounced OMC\%. We note that this argument is based on the relaxation-time approximation: additional extrinsic processes, such as skew scattering, side-jump mechanisms, etc., may introduce new scattering time-dependent odd-$B$ responses~\cite{Nui_prb20, MacDonald_prbR18}. However, such processes do not alter the underlying symmetry constraints or the leading $\tau$-scaling.}

% ~\footnote{We note that this explanation is based on the relaxation-time approximation, where momentum relaxation is dominated by a single effective scattering rate. Additional extrinsic processes, such as skew scattering, side-jump contributions, or strongly energy-dependent impurity potentials, may introduce quantitative modifications. However, these mechanisms do not alter the underlying symmetry constraints or the leading scaling with the TRS-breaking parameter. We therefore expect the qualitative conclusion to remain robust even when more realistic scattering channels are included.}

\section{Crystalline Symmetry restriction on OMC} \label{Sec_3}
In this Section, we analyze the crystalline symmetry restrictions on both the longitudinal and transverse components of the OMC. Since these conductivities represent the linear response in the applied electric field, inversion symmetry ($\mathcal{P}$) does not forbid them. The odd-parity longitudinal magnetoconductivity requires that the TRS of the materials be intrinsically broken even in the absence of the applied magnetic field. While this follows from Onsager's reciprocal relation, below we provide a generic argument for a general magnetoconductivity with arbitrary magnetic field and scattering time dependence. 

Any linear response current can be expressed as
\bea \label{TRS_symm}
j_a =\sigma_{ab} E_b \equiv \tau^p \chi_{ab;c} B_{c}^q E_b~,
\eea 
where $p$ and $q$ denote the power of scattering time and $B$, respectively. Here, we used the Einstein summation convention, where the repeated indices are summed over. $\chi_{ab;c}$ denotes the effective response of a material, which depends only on its intrinsic properties. Under time-reversal operation $\cal T$, $j \to -j$, $\tau \to -\tau$, $B\to -B$ and $E\to E$. Consequently, the time reversed partner of Eq.~\eqref{TRS_symm} becomes $- j_a =  (-1)^{p+q} \tau^p ({\cal T}\chi_{ab;c}) B_{c}^q E_b$, implying 
\be
\chi_{ab;c} \equiv (-1)^{p+q+1} ({\cal T}\chi_{ab;c})~.
\ee 
Thus, $j_a$ can be finite only if the intrinsic response tensor  satisfies $(\mathcal{T}\chi_{ab;c}) = -\chi_{ab;c}$ whenever $p+q$ is even, and $(\mathcal{T}\chi_{ab;c}) = \chi_{ab;c}$ when $p+q$ is odd. Specifically, for the OMC term in Eq.~\eqref{sigma_B}, we have  $p=1,q=1$, and this mandates $\chi_{ab;c}$ to be $\mathcal{T}$-odd for having a finite current. Additionally, a $\mathcal{T}$-odd $\chi_{ab;c}$ can only arise in a magnetic system with intrinsically broken TRS. This explains why the $\tau$-dependent, $B$-linear OMC can be finite only in the magnetic materials.

Beyond these fundamental symmetry constraints, crystallographic point group symmetries further restrict the allowed responses. Since the anomalous Hall conductivity (AHC) can also be finite only in TRS-broken systems, it is natural to ask whether AHC and OMC can coexist under the same symmetry conditions. Now, to systematically determine the crystalline symmetry constraints for OMC, we use the following tensorial form, $\sigma^{\rm OMC}_{ab} = \tau \chi_{ab;c}^{\rm OMC} B_c$. Here, $\chi_{ab;c}^{\rm OMC}$ is the $\mathcal{T}$-odd axial tensor associated with OMC. The anomalous Hall conductivity $\sigma^{\rm AH}_{ab}$ is a polar tensor. In Jahn notation, $\chi^{\rm OMC}_{ab;c}$ maps to $\texttt{ae}[\rm V^2]V$ and $\sigma^{\rm AH}_{ab}$ to $\texttt{a}\{\rm V^2\}$, with $\texttt a$ and $\texttt e$ denoting magnetic and axial tensors; the curly brackets indicate antisymmetry in spatial indices. For a general point group operation ${\cal O}$, the OMC and the AHC response tensors transform as~\cite{newnham_symmetry, Gallego_cryst19}
\begin{subequations}
\bea
\chi_{a'b';c'}^{\rm OMC} &=& \eta_{\cal T} {\rm det}\{ {\cal O} \} {\cal O}_{a'a} {\cal O}_{b'b} {\cal O}_{c'c} \chi_{ab;c}^{\rm OMC}, \\
\sigma_{a'b'}^{\rm AH} &=& \eta_{\cal T}  {\cal O}_{a'a} {\cal O}_{b'b} \sigma_{ab}^{\rm AH}~.
\eea
\end{subequations}
Here, $\eta_{\cal T}=-1$ for magnetic operations ${\cal O}\equiv {\cal RT}$ and $\eta_{\cal T}=+1$ for nonmagnetic operations ${\cal O}\equiv {\cal R}$, with $\cal R$ denoting a pure spatial operation.

The resulting symmetry restrictions for $\chi_{xx;z}^{\rm OMC}$, $\chi_{yx;z}^{\rm OMC}$, and $\sigma_{yx}^{\rm AH}$ are summarized in Table~\ref{table_symmetry}, for common crystalline symmetry elements. Notably, we find that $\chi_{xx;z}^{\rm OMC}$ and $\sigma_{yx}^{\rm AH}$ obey identical symmetry constraints and are allowed under the same point groups, implying that longitudinal OMC and the anomalous Hall effect always coexist. A complete classification of point groups allowing finite $\chi_{xx;z}^{\rm OMC}$ and $\sigma^{\rm AH}_{yx}$ is given in Table~\ref{table_sym_point} of Appendix~\ref{app_B}. {We mention that although $\sigma_{xx}^{\rm OMC}$ and $\sigma^{\rm AH}_{yx}$ obey the same symmetry constraints, they appear in transport measurements in different channels. The OMC modifies the longitudinal conductivity and is odd under $B \to -B$, allowing it to be isolated by antisymmetrizing the longitudinal resistivity. In contrast, the anomalous Hall response contributes to the transverse response and is obtained from the $B = 0$ offset of the Hall resistivity. These two effects, therefore, do not mix in experiments and can be independently extracted.}

Because $\chi_{xx;z}^{\rm OMC}$ is symmetry-allowed under the same conditions as the AHE, detecting OMC provides a direct and complementary probe of intrinsic time-reversal-symmetry breaking. Unlike the AHC, however, the longitudinal OMC manifests in the primary transport channel, making it experimentally accessible even in systems where anomalous Hall voltage or Hall angle is small or masked by extrinsic effects~\cite{Dewai_prb25, Jin_jpsj17, Dulal_sr19, Lewiner73, Zhao_24extrinsicsuppression, chowdhury_24suppression, manna2018colossal}.

The transverse odd-parity response $\chi_{yx;z}^{\rm OMC}$ follows distinct symmetry rules. Specifically, in systems with an in-plane mirror or a twofold in-plane rotation symmetry, the anomalous Hall effect is forbidden, but a symmetric (in spatial indices) transverse Hall response $\propto B$ can still arise. This highlights that the symmetric $B$-linear transverse magnetoresponse also provides an independent and sensitive probe of intrinsic TRS breaking~\cite{Orenstein_23linear}.

 %This distinction underscores the central role of longitudinal OMC as an experimentally robust probe of intrinsic TRS breaking. 

\section{Odd-parity magnetoconductivity in valley-polarized gapped graphene}\label{Sec_4}

\begin{figure}[t]
    \centering
    \includegraphics[width=\linewidth]{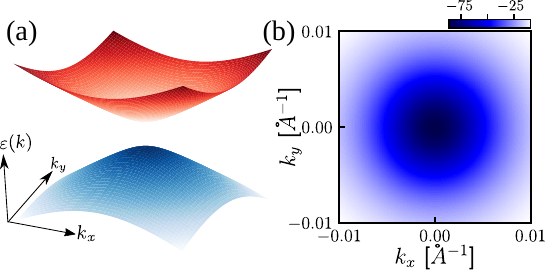}
    \caption{(a) Band dispersion of gapped Dirac Hamiltonian [Eq.~\eqref{Ham_dirac}] for the $\xi=+1$ valley with $v_F=2\times10^5$ m/s, and $\Delta=0.01$ eV. (b) Berry curvature distribution in the $k_x$-$k_y$ plane for the conduction band of the $\xi=+1$ valley, showing pronounced peaks near the band edges.  The orbital magnetic moment shows a similar distribution.} 
    %, with the same sign for both valleys in TRS broken systems.}
    \label{fig2}
\end{figure}
%
% \subsection{Low magnetic field limit}

Having derived the OMC and analyzed its crystalline symmetry constraints, we now calculate it explicitly for the TRS broken gapped Dirac system. We consider a single-layer gapped graphene model and show that the $B$-linear magnetoconductivity vanishes unless the TRS breaks intrinsically. The TRS of graphene can be broken by proximitizing the graphene layer with van-der-Waals magnets such as \ch{Cr2Ge2Te6}, \ch{MnBi2Te4}, or other ferromagnetic materials. The corresponding minimal model Hamiltonian around the $\xi (=\pm 1 )$ valley can be expressed as, 
\be \label{Ham_dirac}
{\cal H}^\xi = \hbar v_F ( \xi k_x \sigma_x + k_y \sigma_y) + \Delta \sigma_z~,
\ee 
where $v_F$ denotes the Fermi velocity and $\Delta$ represents a sublattice potential that generally breaks inversion symmetry. TRS breaking can be introduced by a Haldane mass term $\Delta \to \xi \Delta_0$ (with $\Delta_0$ being the TRS-breaking strength), which generates valley polarization. A similar Hamiltonian also works for the TRS-broken surface states of a topological insulator, with a single Dirac cone~\cite{Zhang_NP09, Zhang_prb10}. With the Haldane mass term, the in-plane mirror symmetries of Hamiltonian~\eqref{Ham_dirac} are broken, although rotational symmetry is preserved~\footnote{For a corresponding tight-binding model Hamiltonian with $\Delta$, the relevant rotation symmetry can be ${\cal C}_{3z}$.}. The energy eigenvalues of Eq.~\eqref{Ham_dirac} are given by $\varepsilon_n({\bm k})= \lambda \sqrt{(\hbar v_F k)^2 + \Delta^2}$, with $ \lambda =\pm$ being the band index. The band dispersion for the $\xi=+1$ valley is shown in Fig.~\ref{fig2}(a). The BC and OMM for the model in Eq.~\eqref{Ham_dirac} are given by $\Omega_z^\xi =- \lambda \xi \Delta \dfrac{\hbar^2 v_F^2 }{2 (\hbar^2v_F^2 k^2 + \Delta^2)^{3/2}}$, and $m_z^\xi =- \xi\Delta \dfrac{e \hbar v_F^2 }{2(\hbar^2v_F^2 k^2 + \Delta^2)}$, respectively. Both the BC and OMM are proportional to $\Delta$, and they vanish for $\Delta=0$, consistent with the restoration of inversion and TRS symmetry for $\Delta=0$. When TRS is broken by introducing $\Delta=\xi \Delta_0$, both the BC and OMM become valley degenerate. These valley degenerate BC and OMM are crucial for generating $\sigma^{\rm OMC}_{xx}(B)$. The BC distribution for the conduction band is shown in Fig.~\ref{fig2}(b), which peaks around the band edges.

\begin{figure*}[t]
    \centering
    \includegraphics[width=\linewidth]{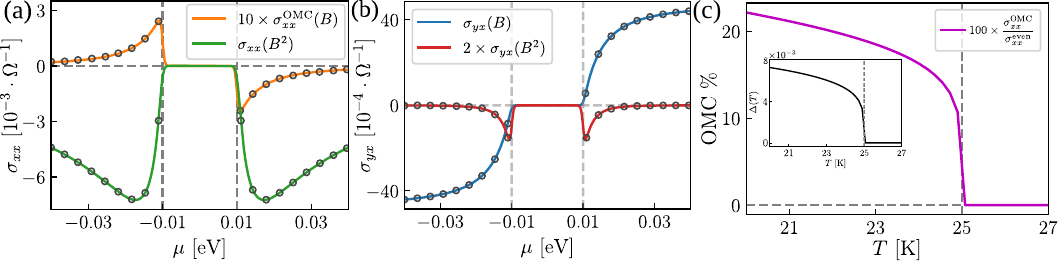}
    \caption{(a) $B$-linear (orange) and the $B$-quadratic (green) longitudinal magnetoconductivities versus chemical potential. (b) $B$-linear conductivity $\sigma_{yx}(B) = \sigma_{yx}^{\rm O}(B) + \sigma_{yx}^{\rm L}(B)$ (blue) and $B$-quadratic conductivity $\sigma_{yx}(B^2)$ (red) versus chemical potential. Both panels are for valley-polarized graphene with $\Delta_0=0.01$ eV, and $B=1$ T. In both (a) and (b), the solid line represents the numerically evaluated result, whereas the open circles denote the analytical results. (c) Numerically calculated temperature dependence of OMC\%, assuming valley polarization of Eq.~\eqref{delta_T_form}. Here, $\sigma^{\rm even}_{xx}$ includes both the Drude and $\sigma_{xx}(B^2)$ contributions. The inset shows the magnetic order parameter $\Delta(T)$ in eV of Eq.~\eqref{delta_T_form} for $\xi = +1$. We have used $\beta=0.20$, $\mu=15$ meV, and $T_c=25$ K. The OMC\% follows $\Delta(T)$ and vanishes at $T_c$, thereby serving as an effective order parameter for TRS breaking in transport experiments.}
    \label{fig3}
\end{figure*}

\subsection{$B$-linear responses} \label{Sec_4A}
Using Eq.~\eqref{sigma_B}, we find that the $B$-linear part of the longitudinal and transverse magnetoconductivities for the Hamiltonian in Eq.~\eqref{Ham_dirac} in the zero-temperature limit is obtained to be
\begin{subequations} \label{sigmas_semi}
\bea 
&&\sigma_{xx}^{\rm OMC}(B)  = - {\rm sgn}(\mu) \xi\Delta\frac{ \tau e^3  v_F^2}{4\pi \hbar} \frac{2\mu^2-\Delta^2}{\mu^4}B, \label{sigmaxx_B_gr} \\
&&\sigma_{yx}^{\rm O}(B)  = -\sigma_{xy}^{\rm O}(B)= - {\rm sgn}(\mu) \frac{ e^3 v_F^2 \hbar}{8\pi}\frac{ \Delta}{\mu^4}B , \\
&&\sigma^{\rm L}_{yx} (B) = -\sigma^{\rm L}_{xy} (B)= {\rm sgn}(\mu) \frac{ \tau^2 e^3  v_F^2}{4\pi \hbar^2} \frac{\mu^2-\Delta^2}{\mu^2}B.~~~~~
\eea
\end{subequations}
Both $\sigma_{xx}^{\rm OMC}(B)$ and $\sigma_{yx}^{\rm O}(B)$ vanish identically for each valley when $\Delta=0$, along with the BC and OMM, as in this case, the Hamiltonian~\eqref{Ham_dirac} preserves both inversion and TRS. Even for $\Delta\neq 0$, the total $\sigma_{xx}^{\rm OMC}(B)$ cancels between valleys, unless valley polarization is introduced via intrinsic TRS breaking captured by $\Delta \to \xi \Delta_0$ in our model. 

The transverse OMC component $\sigma_{yx}^{\rm OMC}(B)$ vanishes due to the rotational symmetry of Hamiltonian~\eqref{Ham_dirac}, see also Table~\ref{table_symmetry}. However, in rotational symmetry–broken magnetic systems, such as magnetized strained graphene, where strain lifts the rotational symmetry, $\sigma_{yx}^{\rm OMC}(B)$ can be finite. In contrast, the transverse response from the intrinsic BC and OMM, $\sigma_{yx}^{\rm O}(B)$, as well as the extrinsic Lorentz Hall effect, $\sigma_{yx}^{\rm L}(B)$, remain finite for \eqref{Ham_dirac}. The intrinsic transverse contribution vanishes for $\Delta=0$, while $\sigma_{yx}^{\rm L}(B)$ persists regardless of TRS or inversion symmetry breaking, consistent with its classical origin from the Lorentz force.

In Figs.~\ref{fig3}(a) and (b), we present the variation of  $\sigma_{xx}^{\rm OMC}(B)$ and $\sigma_{yx}(B)=\sigma_{yx}^{\rm O}(B)+\sigma_{yx}^{\rm L}(B)$ with $\mu$, respectively. Since all the conductivities are Fermi surface effects, they vanish in the band gap region. The $\sigma_{xx}^{\rm OMC}(B)$ originating from band geometric quantities shows sharp peaks near the band edges, reflecting enhanced BC and OMM near the band edges. In contrast, the transverse conductivity $\sigma_{yx}(B)$ grows monotonically with $\mu$, reflecting the dominant role of the classical Lorentz force contribution $\sigma^{\rm L}_{yx}(B)$ over the comparatively weaker band-geometry correction $\sigma^{\rm O}_{yx}(B)$ in $\sigma_{yx}(B)$.

\subsection{$B$-quadratic responses} \label{Sec_4B}
Beyond the $B$-linear contribution, we also find $B$-quadratic longitudinal and transverse magnetoconductivities. Apart from the conventional longitudinal magnetoconductivity $\propto \tau^3 B^2$ arising from the Lorentz force~\cite{AshcroftMermin, Pippard}, we find several other contributions $\propto \tau B^2$, driven by the BC and the OMM. Similarly, we also find various contributions to the transverse magnetoconductivity $\propto B^2$. Interestingly, $\sigma_{yx}(B^2)$ has an intrinsic contribution which is $\tau$-independent along with terms proportional to $\tau, \,  \tau^2$ and $\tau^3$. See Eq.~\eqref{sigma_B2} of Appendix~\ref{app_A} for the detailed expression of $\sigma_{ab}(B^2)$. % of $\sigma_{xx}(B^2)$ and $\sigma_{yx}(B^2)$. 

We calculate $\sigma_{xx}(B^2)$ and $\sigma_{yx}(B^2)$ for the Hamiltonian in Eq.~\eqref{Ham_dirac}, and these are given by 
\begin{widetext}
\begin{subequations} \label{sigmaB2}
\bea \label{sigmaB2_xx}
\sigma_{xx}(B^2)&=& \frac{ e^4  v_F^4 }{4\pi } \left[-\frac{\tau \Delta^2}{4 }\frac{7\mu^2-22\Delta^2}{|\mu|^7}-\frac{\tau^3}{\hbar^2}\frac{\mu^2-\Delta^2}{|\mu|^3}\right]B^2~, \\
\sigma_{yx}(B^2)&=& -\sigma_{xy}(B^2) = -\xi \Delta\frac{ e^4 v_F^4  }{4\pi} \left[ \frac{3 \hbar \Delta^2}{4 }\frac{1}{|\mu|^7} + \frac{\tau^2}{\hbar} \frac{2 \mu^2 - \Delta^2}{|\mu|^5} \right]B^2~.  \label{sigmaB2_yx}
\eea
\end{subequations}
\end{widetext}
The first term of $\sigma_{xx}(B^2)$ in Eq.~\eqref{sigmaB2_xx} originates from BC and OMM, while the second term arises from the conventional Lorentz force~\cite{Pippard}. Both are $\mathcal{T}$-even, hence, can be finite in magnetic as well as nonmagnetic materials (see also Sec.~\ref{Sec_3} and Appendix~\ref{app_B}). In contrast, the $\mathcal{T}$-odd quadratic components ($\propto \tau^0$ and $\tau^2$) do not contribute to $\sigma_{xx}(B^2)$, since they are antisymmetric in spatial indices and therefore represent genuine Hall-type responses.

For the transverse conductivity, however, the situation differs. For the model in Eq.~\eqref{Ham_dirac}, only the $\tau^0$ and $\tau^2$ terms remain finite and are valley dependent. The responses in Eq.~\eqref{sigmaB2_yx} are antisymmetric in spatial indices, representing a genuine Hall response. On the other hand, the rotational symmetry of Hamiltonian~\eqref{Ham_dirac} enforces the $\mathcal{T}$-even components $\propto \tau$ and $\tau^3$ in $\sigma_{yx}(B^2)$ to vanish (see Table~\ref{table_symmetry2} and Appendix~\ref{app_A}). Importantly, in systems where rotational symmetry is broken, they can, in general, become finite. This would generate a spatially symmetric $B$-quadratic transverse response even in TRS-preserved systems.

We plot $\sigma_{xx}(B^2)$ and $\sigma_{yx}(B^2)$ of Eqs.~\eqref{sigmaB2} as a function of $\mu$ in Fig.~\ref{fig3}(a) and (b). Classically, the longitudinal conductivity is $\propto B^2$, and the transverse conductivity is $\propto B$. From Fig.~\ref{fig3}(a-b), we observe that the dominant responses are these classical contributions. The band-geometry driven contributions to $\sigma^{\rm OMC}_{xx}(B)$ and $\sigma_{yx}(B^2)$ are orders of magnitude smaller than the classical contribution and only show prominent responses near the band edges.

\subsection{OMC as a transport order parameter for TRS breaking} \label{Sec_4C}
We now examine how the longitudinal odd-parity conductivity $\sigma_{xx}^{\rm OMC}$ depends on temperature, emphasizing its role as a probe of intrinsic TRS breaking. From Eq.~\eqref{sigmaxx_B_gr}, the $\sigma_{xx}^{\rm OMC}(B)$ is valley dependent and proportional to $\xi \Delta$. On the other hand, the $\cal T$-even components such as $\sigma_{xx}(B^2)$ and $\sigma_{yx}(B)$ are valley independent. Thus, $\cal T$-odd components become finite only in valley-polarized TRS broken graphene ($\Delta \to \xi \Delta_0$). As the temperature approaches the critical temperature $T_c$ of the graphene–magnetic material heterostructure, the magnetization vanishes and the intrinsic TRS is restored. Consequently, {$\sigma_{xx}^{\rm OMC}(B)$} should also vanish above $T_c$.

To study this behavior, we assume that the TRS-breaking magnetic order follows the second-order phase transition, 
\bea \label{delta_T_form}
\Delta(T) = \begin{cases}
\xi \Delta_0 (1 - T/T_c)^\beta &~\text{for}~~ T < T_c, \\
0 &~\text{for}~ ~T \geq T_c.
\end{cases}
\eea
Here, $\beta$ is the critical exponent, which typically ranges from $0.16$ to $0.46$ for two-dimensional van-der-Waals magnetic materials~\cite{Wang_acsn22}. 
\begin{figure*}[t]
    \centering
    \includegraphics[width=\linewidth]{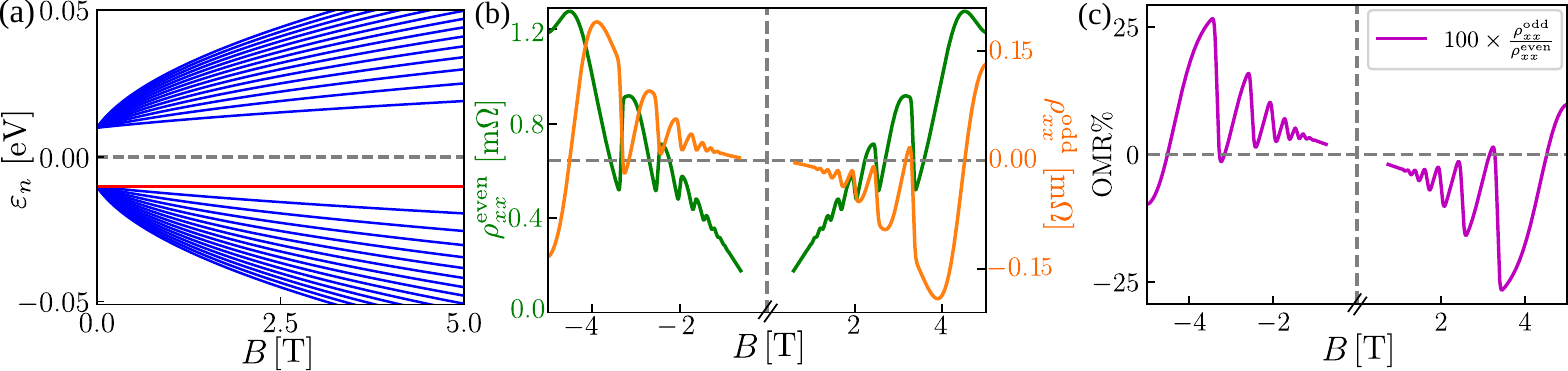}
    \caption{(a) Landau levels of Hamiltonian~\eqref{Ham_B} for the $\xi=+1$ valley. The zeroth Landau level lies at $-\Delta$ for this valley, and it is independent of $B$. (b) Even-parity ($\rho^{\rm even}_{xx}$) and odd-parity ($\rho_{xx}^{\rm odd}$) resistivities for valley polarized gapped Dirac fermions, showing the symmetric and antisymmetric dependence on $B$. (c) Odd-parity magnetoresistivity (OMR\%) as a function of $B$ for chemical potential $\mu=0.025$ eV. We have taken $k_s=10^8~\rm m^{-1}$, $n_{\rm im}=10^{13}~ \rm m^{-2}$, $\Gamma_0=1$ meV, and $\Delta=0.01$ eV.}
    \label{fig4}
\end{figure*}

% the $\Delta=\Delta_0$ only breaks the inversion symmetry of Hamiltonian~\eqref{Ham_dirac}, crucial for finite band geometric contributions to the $\cal T$-even conductivities.

Using Eq.~\eqref{delta_T_form} and assuming a representative $T_c=25$ K, we numerically evaluate the longitudinal magnetoconductivities as a function of temperature and plot the OMC\% following Eq.~\eqref{OMC_per}. {From Fig.~\ref{fig3}(c),} it is clear that the $\rm OMC\%$ vanishes as temperature approaches $T_c$. Remarkably, OMC\% follows the behavior of $\Delta(T)$ vs. $T$ shown in the inset of Fig.~\ref{fig3}(c). This is because the OMC\%, to leading order, is directly proportional to $\Delta(T)$, see Appendix~\ref{app_C_new} for details. Thus, the $\rm OMC\%$ effectively acts as a transport order parameter for probing intrinsic TRS-breaking in a material. A similar observation was reported in Ref.~\cite{Sahani_prl24}, where the odd-parity magnetoresistivity {[hence $\sigma_{xx}^{\rm OMC}(B)$]} was observed to vanish beyond a certain temperature in the graphene-\ch{Cr2Ge2Te6} sample. We emphasize that scattering times can also vary with temperature in real materials; here, we employ the constant relaxation-time approximation to qualitatively highlight the critical behavior of OMC\%. 

\section{OMC in quantizing magnetic field } \label{Sec_5}

In the previous section, we analyzed the magnetoconductivities in the semiclassical low-field regime, where Landau levels are not resolved. However, the fate of odd-$B$ magnetoconductivities in the quantum oscillation regime remains unexplored. Here, we explicitly evaluate the magnetoconductivity (and resistivity) in the presence of a quantizing magnetic field, where multiple discrete Landau levels contribute to transport.

We incorporate the magnetic field in Eq.~\eqref{Ham_dirac} via minimal coupling ${\hbar \bm k} \to (\hbar {\bm k} +e{\bm A})$, with ${\bm B}={\bm \nabla}_{\bm r} \times {\bm A}$. Consequently, the Hamiltonian~\eqref{Ham_dirac} becomes 
\be \label{Ham_B}
{\cal H}_B^\xi =  v_F \big[ \xi(\hbar k_x +eA_x) \sigma_x + (\hbar k_y +eA_y) \sigma_y) \big] + \Delta \sigma_z.
\ee
To evaluate the Landau levels in the presence of an out-of-plane magnetic field, we choose the Landau gauge, ${\bm A} =  (0,xB,0)$. The Landau levels (LL) are obtained to be, 
\bea \label{LLs}
\varepsilon_n = \begin{cases}\lambda \sqrt{\Delta^2 + 2n (\hbar \omega_c)^2 } & n\neq 0, \\
-\xi \Delta & n=0.
\end{cases}
\eea
Here, $\lambda = \pm 1$ represents the band index for conduction and valence bands, $n$ denotes the Landau levels indices, and $\omega_c = v_F/l_B $ is the cyclotron frequency with $l_B = \sqrt{\hbar / eB}$ being the magnetic length. The corresponding wave functions can be represented in terms of the Harmonic oscillator eigenstates, as detailed in Appendix~\ref{app_C}. Note that in the absence of a sublattice potential, the zeroth LL lies at zero energy and is valley-degenerate. A finite $\Delta$ breaks the valley degeneracy of the zeroth LL, where it has energy of $\varepsilon_0=  -\Delta$ ($\varepsilon_0= \Delta$) for the $\xi =+1$ ($\xi =-1$) valley. In Fig.~\ref{fig4}(a), we show the LLs for the $\xi = +1$ valley. The non-dispersing zeroth LL is marked by the red horizontal line.

In an ideal 2D system with momentum-independent, perfectly flat Landau bands, the group velocity vanishes, and longitudinal conductivity would be zero. However, in real
materials, impurities and disorder broaden these levels, creating extended states where electrons can scatter between
Landau orbits. This scattering, driven by impurities, gives rise to collisional conductivity, which governs
longitudinal magnetoconductivity at low temperatures. This collisional conductivity can be evaluated using the following formula~\cite{Vasil_jmp82, Vasil_prb92,  islam2018magnetotransport},
\bea \label{coll_conduct}
    \sigma^{\text{col}}_{xx}(B) = \frac{\beta_T e^2}{2 S_0} \sum_{\zeta, \zeta'} f_\zeta^0 \left(1 - f_{\zeta'}^0 \right) W_{\zeta \zeta'} \left(x_\zeta - x_{\zeta'}\right)^2.
\eea
Here, the $\zeta$ includes all the quantum index $\{\xi,\lambda,n,k_y\}$, $S_0$ is the sample area and $\beta_T=1/k_BT$ is the inverse temperature. %$f_\zeta^0 = f_{\zeta'}^0$ for elastic scattering. 
The $W_{\zeta \zeta'}$ represents the scattering rate between states $\zeta$ and $\zeta'$. The $x_\zeta = \bra{\zeta} x \ket{\zeta} = -k_y l_B^2$ is the average value of the $x$-component of the position operator of an electron in a particular quantum state. For the elastic scattering and static impurities, the scattering rate between states $\zeta$ and $\zeta'$ is given by,
\be \label{scat rate}
    W_{\zeta \zeta'} = \frac{2\pi n_{\rm im}}{S_0 \hbar} \sum_q |U_q|^2 |F_{\zeta \zeta'}(\eta)|^2 \delta(\varepsilon_\zeta - \varepsilon_{\zeta'}),
\ee
where $n_{\rm{im}}$ is the impurity density. The Fourier transformation of the screened charged impurity potential $U(r) = \left[\frac{e^2}{4\pi \epsilon_0 \epsilon_r r}\right] e^{-k_s r}$ is given by $U_q = U_0 [q^2 + k_s^2]^{-1/2} \approx \frac{U_0}{k_s}$ under the limit of small $ |q| \ll k_s $ for short-range Delta function-like potential. Here, $r$ is the real space distance from the impurity, $\epsilon_0$ is free space permittivity, and $\epsilon_r$ denotes relative permittivity. Here, $U_0 = \frac{e^2}{2 \epsilon_0 \epsilon_r}$ and $k_s$ is the screening vector. The function $F_{\zeta \zeta'}(\eta)=\bra{\zeta}  e^{i{\bm q}\cdot{\bm r}}\ket{\zeta'}$ denotes the form factor with its argument being $\eta = q^2 l_B^2 / 2$.  

% \textcolor{red}{what is $\epsilon_0, etc$}

After a detailed calculation (see Appendix~\ref{app_C} for details), we obtain the collisional magnetoconductivity in the zero temperature limit to be
\be \label{sigma_osc}
\sigma_{xx}^{\rm col}(B) = \sigma_0(B)  \sum_{n,\xi} \left[ 2n\left(1+3\frac{ \Delta^2 }{\varepsilon_n^2}\right)  - \xi \frac{4\Delta}{\varepsilon_n} \right] \delta(\varepsilon_n- \mu),
\ee
with $\sigma_0(B)=\dfrac{e^2}{h}\dfrac{n_{\rm im}U_0^2}{2\pi k_s^2 l_B^2\Gamma_0}$. Equation~\eqref{sigma_osc} captures the oscillatory contribution from the $n\!\neq\!0$ Landau levels. Unlike the low-field semiclassical regime, where conductivities scale with definite powers of $B$, $\sigma_{xx}^{\rm col}(B)$ in the quantum oscillation regime contains terms at all orders in $B$. It is therefore natural to analyze the results by separating even and odd components under $B\to - B$. We find that in Eq.~\eqref{sigma_osc}, the term linear in $\Delta$ is odd in $B$ and proportional to the valley index $\xi$. Consequently, the corresponding charge transport vanishes in valley-degenerate systems but becomes finite when valleys are polarized (systems with intrinsic TRS breaking). 
In contrast, the terms proportional to $2n$ in Eq.~\eqref{sigma_osc} are even in $B$ and survive regardless of valley polarization. Thus, an intrinsically TRS-broken system exhibits both even and odd oscillatory components in the charge transport, consistent with the semiclassical results.

The contribution of the zeroth LL to the conductivity is 
\be \label{sigma_ultra}
\sigma_{xx}^{\rm col}(B) = \sigma_0(B) \sum_\xi \delta(\xi \Delta + \mu)~.
\ee
We find that this contribution is always $B$-linear and remains finite even when $\xi\Delta \to 0$. Hence, the zeroth LL yields a universal $B$-linear magnetoconductivity in graphene, regardless of whether time-reversal or inversion symmetry is broken. This behavior is consistent with Abrikosov's quantum theory of linear magnetoresistance~\cite{Abrikosov_prb98, Abrikosov2000}, which describes systems with linear band dispersion under strong magnetic fields. In the ultra-quantum regime, where only the lowest Landau level is occupied, the magnetoresistance is therefore universally linear in $B$, irrespective of intrinsic TRS breaking, in systems with relativistic band dispersion. 

Using the odd-$B$ and even-$B$ components of $\sigma_{xx}^{\rm col}(B)$ from Eqs.~\eqref{sigma_osc} and \eqref{sigma_ultra}, the corresponding odd and even-parity magnetoresistivities can be obtained using the relation $\rho_{xx} = \dfrac{\sigma_{xx}}{\sigma_{xx}^2 + \sigma_{xy}^2}$,
with $\sigma_{xy} = 4 \dfrac{e^2}{h}(n+\tfrac{1}{2})$ being the quantized Hall conductivity~\cite{Novoselov_nat05, Zhang_nat05}. Similar to the OMC\% in Eq.~\eqref{OMC_per}, we charactarize the odd-parity magnetoresistivity (OMR) as $
{\rm OMR}\% = \frac{\rho^{\rm odd}_{xx}}{\rho^{\rm even}_{xx}} \times 100 $. Figure~\ref{fig4}(b) shows $\rho_{xx}^{\rm even}$ and $\rho_{xx}^{\rm odd}$ versus $B$ for a representative chemical potential $\mu=0.025$ eV. As expected, $\rho_{xx}^{\rm even}$ is symmetric in $B$, while $\rho_{xx}^{\rm odd}$ is antisymmetric and the odd component is roughly an order of magnitude smaller. In Fig.~\ref{fig4}(c), we plot OMR\% as a function of $B$ for the same $\mu$. The OMR\% increases with the field strength, reaching $\sim 26\%$ at $B\!\sim\!4$ T. Similar behavior has been observed experimentally in graphene proximitized by the ferromagnetic insulator \ch{Cr2Ge2Te6}, where magnetic proximity breaks intrinsic TRS and generates odd-parity magnetoresistance across both low- and high-field regimes~\cite{Sahani_prl24}.

\section{Conclusion} \label{Sec_6}
% \textcolor{red}{make it shorter and less repetitive}

In summary, we have shown that odd-parity magnetoconductivity (OMC) emerges as a distinct and robust transport signature of intrinsic time-reversal symmetry breaking in magnetic metals. Unlike conventional magnetoconductivity, which is even in $B$, OMC varies linearly with $B$ and originates from band-geometric quantities such as Berry curvature and orbital magnetic moment. Our detailed crystalline symmetry analysis reveals that longitudinal OMC shares the same constraints as the anomalous Hall effect, while the transverse OMC follows distinct rules. We explicitly calculate the conductivities for valley-polarized gapped graphene and show that the response peaks near the band edges and vanish in the band gap, reflecting their Fermi surface origin. Furthermore, our numerical calculations for the temperature dependence of OMC show that it directly follows the underlying TRS-breaking order parameter. Thus, OMC effectively behaves as a TRS breaking order parameter in transport experiments. 

Beyond the $B$-linear responses, we identified quadratic in $B$ conductivities that combine conventional Lorentz contributions with additional band-geometric terms, each exhibiting distinct scattering-time scalings. In particular, alongside the conventional $\mathcal{T}$-even longitudinal magnetoconductivity $\propto \tau^3 B^2$ from the Lorentz force, Berry curvature and orbital magnetic moment generate additional longitudinal contributions $\propto \tau B^2$. For the transverse response, we uncover both intrinsic $\tau^0 B^2$ and extrinsic $\tau^2 B^2$ terms, which appear only in magnetic metals. Additionally, in rotational symmetry broken materials, transverse magnetoconductivity $\propto \tau B^2$ and $\tau^3 B^2$ can also appear.

Finally, we evaluated the magnetoconductivity and resistivity for proximity magnetized graphene in the quantum oscillation regime with distinguishable Landau levels. We find that the oscillating magnetoconductivity contains both even- and odd-in-$B$ components, with the zeroth Landau level producing a universal $B$-linear contribution consistent with Abrikosov’s quantum theory. The resulting oscillating resistivity with $B$ naturally explains the experimentally observed odd-parity magnetoresistance in Ref.~\cite{Sahani_prl24}. Our results establish OMC and its resistive counterpart (OMR) as powerful probes of intrinsic time-reversal symmetry breaking, complementary to anomalous Hall transport, and relevant for identifying topological magnetic phases. 

\section{Acknowledgement}
We thank Sayan Sarkar (IIT Kanpur) for help with the crystallographic symmetry analysis. S. Das acknowledges the Ministry of Education, Government of India, for funding support through the Prime Minister's Research Fellowship. A. Adhikary is supported by the Institute Fellowship, IIT Kanpur. A. Agarwal acknowledges funding from the Core Research Grant by ANRF (Sanction No. CRG/2023/007003), Department of Science and Technology, India.

% \clearpage
\onecolumngrid

\appendix
% \begin{widetext}
\section{Derivation of magnetoconductivity expressions } \label{app_A}
In this Appendix, we calculate magnetoconductivity expressions up to second order in magnetic field in the linear response regime, using the expressions of the $\dot{\bm r}$ and $g_{n\bm{k}}$. Since we are interested up to second order in the magnetic field, we take up to $l=2$ in Eq.~\eqref{distr_fn}. Semiclassically, current density can be expressed as,
\begin{eqnarray}
    \bm{j}&&=-e \sum_n \int [d\bm{k}] {\cal D}^{-1}\dot{\bm{r}} g_{n\bm k} \nonumber \\
    &&= -e \sum_n \int [d\bm{k}] \left( \bm{\tilde v} +  \frac{e}{\hbar} \bm{E}\times \bm{\Omega} \right)\nn\\
    &&\quad \left[ f_n^0 + e\tau {\cal D} \left( \bm{\tilde v}\cdot \bm{E}\right)\frac{\partial f_n^0}{\partial \varepsilon} + \left( \frac{e\tau}{\hbar} {\cal D}(\bm{\tilde v} \times \bm{B})\cdot\bm\nabla_{\bm{k}} \right) \left[e \tau {\cal D} ( \bm{\tilde v} \cdot \bm{E} ) \frac{\partial f_n^0}{\partial \varepsilon} \right]+\left( \frac{e\tau}{\hbar} {\cal D}(\bm{\tilde v} \times \bm{B})\cdot\bm\nabla_{\bm{k}} \right)^2 \left[e \tau {\cal D} ( \bm{\tilde v} \cdot \bm{E} ) \frac{\partial f_n^0}{\partial \varepsilon} \right]\right]\nonumber \\ 
    && = -e\int_{n,{\bm k}} \bm{\tilde v} f_n^0 -\frac{e^2}{\hbar}\int_{n,{\bm k}} (\bm{E}\times \bm{\Omega})f_n^0 - e^2 \tau\int_{n,{\bm k}} {\cal D}{(\bm{\tilde v} \cdot \bm{E})\bm{\tilde v}\frac{\partial f_n^0}{\partial \varepsilon}}\nonumber \quad- \frac{e^3 \tau^2}{\hbar}\int_{n,{\bm k}}\bm{\tilde v} {\cal D}(\bm{\tilde v} \times \bm{B})\cdot\bm\nabla_{\bm{k}}\left[{\cal D} ( \bm{\tilde v} \cdot \bm{E} )\frac{\partial f_n^0}{\partial \varepsilon}\right]\nn\\
    && \quad -\frac{e^4 \tau^3}{\hbar^2}\int_{n,{\bm k}}\bm{\tilde v} \left({\cal D}(\bm{\tilde v} \times \bm{B})\cdot\bm\nabla_{\bm{k}}\right)^2\left[{\cal D} ( \bm{\tilde v} \cdot \bm{E} )\frac{\partial f_n^0}{\partial \varepsilon}\right],
\end{eqnarray}
For brevity, we have used the notation $\int_{n,{\bm k}} \equiv \sum_n \int \frac{d^d k}{(2\pi)^d}$ for $d$-dimensional system. In the above equation, the first term is zero, representing the equilibrium current. The second term is the anomalous Hall effect. To explicitly derive the magnetoconductivities, we use the relations ${\cal D}=(1 + \frac{e}{\hbar}{\bm\Omega} \cdot {\bm B})^{-1} \approx 1- \frac{e}{\hbar}{\bm\Omega} \cdot {\bm B} + \frac{e^2}{\hbar^2}({\bm\Omega} \cdot {\bm B})^2 +\cdots $, $f_n^0(\tilde \varepsilon) \approx f_n^0(\varepsilon) - {\bm m}\cdot{\bm B} (\partial_{\varepsilon} f_n^0) + \frac{{\left({\bm m}\cdot{\bm B}\right)}^2}{2}(~\partial^2_{\varepsilon} f_n^0)+\cdots$. Using these, the current density up to second order in the magnetic field can be derived from the following expression 
\bea 
\bm j &=& -\frac{e^2}{\hbar}\int_{n,{\bm k}} (\bm{E}\times \bm{\Omega})\Big( f_n^0
- {\bm m}\cdot{\bm B} (\partial_{\varepsilon} f_n^0) 
+ \frac{\left({\bm m}\cdot{\bm B}\right)^2}{2} (\partial^2_{\varepsilon} f_n^0) 
 \Big) \nn\\
&& -e^2 \tau\int_{n,{\bm k}} \Big[ \left(1 - \frac{e}{\hbar}{\bm \Omega} \cdot {\bm B} + \frac{e^2}{\hbar^2}({\bm \Omega} \cdot {\bm B})^2 \right) \times ({\bm v}^0 + {\bm v}^m)\Big(({\bm v}^0 + {\bm v}^m) \cdot \bm{E}\Big)  \Big] \nn\\
&& ~~~~ \times \frac{\partial}{\partial \varepsilon} \Big[ f_n^0
- {\bm m}\cdot{\bm B} (\partial_{\varepsilon} f_n^0) 
+ \frac{\left({\bm m}\cdot{\bm B}\right)^2}{2} (\partial^2_{\varepsilon} f_n^0) 
\Big]\nn\\
&&- \frac{e^3 \tau^2}{\hbar}\int_{n,{\bm k}}({\bm v}^0 + {\bm v}^m) \left(1 - \frac{e}{\hbar}{\bm \Omega} \cdot {\bm B} \right)(({\bm v}^0 + {\bm v}^m) \times \bm{B})\cdot\bm\nabla_{\bm{k}}\left[\left(1 - \frac{e}{\hbar}{\bm \Omega} \cdot {\bm B} \right) ( ({\bm v}^0 + {\bm v}^m) \cdot \bm{E} )\right]\nn\\
&& ~~~~ \times \frac{\partial}{\partial \varepsilon} \Big[ f_n^0
- {\bm m}\cdot{\bm B} (\partial_{\varepsilon} f_n^0) \Big]-\frac{e^4 \tau^3}{\hbar^2}\int_{n,{\bm k}}\bm{v^0} \left((\bm{v^0} \times \bm{B})\cdot\bm\nabla_{\bm{k}}\right)^2\left[ ( \bm{v^0} \cdot \bm{E} )\frac{\partial f_n^0}{\partial \varepsilon}\right].
\eea
Note that we don't have any terms containing $(\bm{\tilde v}\cdot{\bm \Omega})$, since for a two-dimensional system it is identically zero. The conductivity expressions corresponding to the current density $j_a$ for the applied electric field $E_b$ are given by ($\epsilon_{abc}$ is the anti-symmetric Levi-Civita tensor)
\bea 
\sigma_{ab}(B^0) &=& -e^2 \tau \int_{n,{\bm k}} {v}_{a}^0 {v}_{b}^0 ~\partial_{\varepsilon} f_n^0 - \frac{e^2}{\hbar} \epsilon_{abc} \int_{n,{\bm k}} {\Omega}_c f_n^0 \label{dr} , \\
\sigma_{ab}(B) &=& \frac{e^2}{\hbar} \epsilon_{abc} \int_{n,{\bm k}}  {\Omega}_c ({\bm m}\cdot {\bm B}) \partial_{\varepsilon} f_n^0 - \frac{e^3 \tau^2}{\hbar} \int_{n,{\bm k}}  { v}_{a}^0 ({\bm v}^0 \times {\bm B}) \cdot {\bm \nabla}_{\bm k} [({ v}_{b}^0)~ \partial_{\varepsilon} f_n^0] \nn  \\
&& - e^2 \tau \int_{n,{\bm k}} \left[ {v}_{a}^m {v}_{b}^0 + { v}_{a}^0  v_{b}^m \right] ~\partial_{\varepsilon} f_n^0 + \frac{e^3\tau}{\hbar} \int_{n,{\bm k}} ({\bm \Omega}\cdot {\bm B}) {v}_{a}^0 { v}_{b}^0 ~\partial_{\varepsilon} f_n^0  + e^2 \tau \int_{n,{\bm k}} { v}_{a}^0 {v}_{b}^0~ \partial_{\varepsilon}\left[ ({\bm m}\cdot {\bm B}) ~\partial_{\varepsilon} f_n^0 \right], \\
\sigma_{ab}(B^2)&=& -\frac{e^2}{2\hbar} \epsilon_{abc} \int_{n,{\bm k}}  {\Omega}_c ({\bm m}\cdot {\bm B})^2 \partial^2_{\varepsilon} f_n^0 -e^2 \tau \int_{n,{\bm k}} { v}_{a}^m {v}_{b}^m ~\partial_{\varepsilon} f_n^0 - \frac{e^4 \tau}{\hbar^2} \int_{n,{\bm k}} ({\bm \Omega}\cdot {\bm B})^2 {v}_{a}^0 {v}_{b}^0 ~\partial_{\varepsilon} f_n^0 \nn \\
&& + \frac{e^3 \tau}{\hbar} \int_{n,{\bm k}} ({\bm \Omega}\cdot {\bm B}) \left[ { v}_{a}^m { v}_{b}^0 + {v}_{a}^0 { v}_{b}^m  \right] ~\partial_{\varepsilon} f_n^0 + e^2 \tau \int_{n,{\bm k}}\left[ { v}_{a}^m { v}_{b}^0  + { v}_{a}^0 { v}_{b}^m  \right] \partial_{\varepsilon}\left[ ({\bm m}\cdot {\bm B}) ~\partial_{\varepsilon} f_n^0 \right] \nn \\
&& +\frac{e^2 \tau}{2} \int_{n,{\bm k}} { v}_{a}^0 {v}_{b}^0 \partial_{\varepsilon}\left[ ({\bm m}\cdot {\bm B})^2 ~\partial^2_{\varepsilon} f_n^0 \right] + \frac{e^3 \tau}{\hbar} \int_{n,{\bm k}} ({\bm \Omega}\cdot {\bm B}) {v}_{a}^0 { v}_{b}^0   \partial_{\varepsilon}\left[ ({\bm m}\cdot {\bm B}) ~\partial_{\varepsilon} f_n^0 \right]\nn \\
&& - \frac{e^3 \tau^2}{\hbar}\int_{n,{\bm k}} { v}_{a}^0 ({\bm v}^0 \times \bm{B})\cdot\bm\nabla_{\bm{k}}({v}_{b}^m ~\partial_{\varepsilon} f_n^0)- \frac{e^3 \tau^2}{\hbar}\int_{n,{\bm k}}\left[{ v}_{a}^0 ({\bm v}^m \times \bm{B})+{ v}_{a}^m ({\bm v}^0 \times \bm{B})\right]\cdot\bm\nabla_{\bm{k}}({ v}_{b}^0 ~\partial_{\varepsilon} f_n^0)\nn\\
&& +\frac{e^4 \tau^2}{\hbar^2}\int_{n,{\bm k}} { v}_{a}^0 ({\bm v}^0 \times \bm{B})\cdot\left[\bm\nabla_{\bm{k}}(({\bm \Omega}\cdot {\bm B}) { v}_{b}^0 )+({\bm \Omega}\cdot {\bm B})\bm\nabla_{\bm{k}} { v}_{b}^0 \right]~\partial_{\varepsilon} f_n^0\nn\\
&& +\frac{e^3 \tau^2}{\hbar}\int_{n,{\bm k}}{ v}_{a}^0 ({\bm v}^0 \times \bm{B})\cdot\bm\nabla_{\bm{k}} \left[{ v}_{b}^0 \partial_{\varepsilon}[({\bm m}\cdot {\bm B}) ~\partial_{\varepsilon} f_n^0]\right]-\frac{e^4 \tau^3}{\hbar^2}\int_{n,{\bm k}}v_{a}^0 \left(({\bm v}^0 \times \bm{B})\cdot {\bm\nabla}_{\bm{k}}\right)^2\left[ v_{b}^0  \partial_{\varepsilon} f_n^0\right].\label{sigma_B2}
\eea
In Eq.~\eqref{dr}, the first term is the Drude conductivity, whereas the second term represents the Berry curvature-induced anomalous Hall conductivity. In the conductivity expressions, the terms with Levi-Civita tensor and $\propto ({\bm v}^0 \times \bm{B})$ only contribute to the transverse conductivity, as they are antisymmetric under the exchange of $a$ and $b$. All other terms, including the term $\propto ({\bm v}^0 \times \bm{B})^2$ contribute to both longitudinal and transverse magnetoconductivities.

\section{Point groups allowing for OMC and symmetry analysis of $\sigma_{ab}(B^2)$} \label{app_B}
\begin{table}[h] 
    \centering
    \caption{The crystalline point groups allowing for finite $\chi_{xx;z}^{\rm OMC}$, $\sigma^{\rm AH}_{yx}$, and $\chi_{yx;z}^{\rm OMC}$.  }
    \renewcommand{\arraystretch}{1}
    \setlength\tabcolsep{0.24 cm}
    \begin{tabular}{c |c }
       \hline \hline
          $\chi_{xx;z}^{\rm OMC}$  &  
        $ 1, -1, 2^{\prime}, m^{\prime}, 2^{\prime}/m^{\prime}, 2^{\prime}2^{\prime}2, m^{\prime} m^{\prime}2,
 mm^{\prime} m^{\prime}, 
4, 4^{\prime}, -4, -4^{\prime}, 4/m, 4^{\prime}/m,  42^{\prime}2^{\prime},4m^{\prime} m^{\prime}, 4^{\prime}m ^{\prime} m,$ \\ & $ -4 2^{\prime}m^{\prime}, -4^{\prime}2^{\prime}m,    4/m m^{\prime}m^{\prime}, 3, -3, 32^{\prime}, 3m^{\prime}, -3m^{\prime}, 6, -6, 6/m, 62^{\prime}2^{\prime}2, 6m^{\prime}m^{\prime}, -6m^{\prime}2^{\prime} $\\
\hline
$\sigma^{\rm AH}_{yx}$ &  $1, -1, 2^{\prime}, m^{\prime}, 2^{\prime}/m^{\prime}, 2^{\prime}2^{\prime}2,  m^{\prime}m^{\prime}2, mm^{\prime}m^{\prime}, 4, -4, 4/m, 42^{\prime}2^{\prime}, 4m^{\prime}m^{\prime}, -4 2^{\prime}m^{\prime}, 4/mm^{\prime}m^{\prime},
3,$  \\ & $ -3, 32^{\prime},  3m^{\prime}, -3m^{\prime}, 6, -6, 6/m, 62^{\prime}2^{\prime}, 6m^{\prime}m^{\prime}, -6m^{\prime}2^{\prime}
$ \\
\hline
$\chi_{yx;z}^{\rm OMC}$ & $
1, -1, 2, m, 2/m, 222, mm2, mmm, 4^{\prime}, -4^{\prime}, 4^{\prime}/m, 4^{\prime} 2 2^{\prime}, -4^{\prime} 2 m^{\prime}, 4^{\prime}/m mm^{\prime}, 23, m3, 4^{\prime}32^{\prime}$
 \\
    \hline \hline
    \end{tabular}
    \label{table_sym_point}
\end{table}

In Table~\ref{table_sym_point}, we list all magnetic point groups under which finite $\chi_{xx;z}^{\rm OMC}$, $\chi_{yx;z}^{\rm OMC}$, and $\sigma_{yx}^{\rm AH}$ are symmetry-allowed. Notably, the point groups $-4^{\prime}, -4^{\prime}2^{\prime}m, 4^{\prime}, 4^{\prime}/m,$ and $4^{\prime}m^{\prime}m$ permit a finite $\chi_{xx;z}^{\rm OMC}$ by symmetry, but forbid $\sigma^{\rm AH}_{yx}$. Importantly, $\chi_{xx;z}^{\rm OMC}$ originates from Berry curvature (BC) and orbital magnetic moment (OMM). Both quantities vanish under the combined symmetry ${\cal C}_{4z}{\cal T}$, which is present in all of these point groups. Thus, despite being symmetry-allowed at the tensor level, a microscopic evaluation yields zero $\chi_{xx;z}^{\rm OMC}$. This establishes that longitudinal OMC and the anomalous Hall conductivity (AHC) always appear together.

% In Table~\ref{table_sym_point}, we list all the magnetic point groups under which finite $\chi_{xx;z}^{\rm OMC}$, $\chi_{yx;z}^{\rm OMC}$ and $\sigma_{yx}^{\rm AH}$ are allowed. From Table~\ref{table_sym_point}, we note that the point groups $-4^{\prime}, -4^{\prime}2^{\prime}m, 4^{\prime}, 4^{\prime}/m, 4^{\prime}m^{\prime}m$ allows finite $\chi_{xx;z}^{\rm OMC}$, however, they don't allow finite $\sigma^{\rm AH}_{yx}$. Crucially, $\chi_{xx;z}^{\rm OMC}$ responses arise from the BC and the OMM. Both the BC and OMM vanish under the symmetry ${\cal C}_{4z}\cal T$, and the above-mentioned point groups all contain ${\cal C}_{4z}\cal T$ as one of the symmetry elements. Consequently, these point groups do not allow for the finite $\chi_{xx;z}^{\rm OMC}$ when the microscopic expression of OMC is taken into account. This concludes that OMC and AHC always appear together.

Now we discuss the crystalline symmetry restriction on the $B$-quadratic magnetoconductivity $\sigma_{ab}(B^2)$. As discussed in the previous section, only $\tau$ and $\tau^3$ terms contribute to the longitudinal conductivity $\sigma_{aa}(B^2)$. Consequently, $\sigma_{aa}(B^2)$ can be finite even in the time-reversal symmetric materials. This can be understood as follows. We can write the corresponding longitudinal current density as 
\be
j_a = \tau^p \chi_{aa;cd} E_a B_c B_d,
\ee 
with $p=1,3$. Here, we have $q=2$, following Eq.~\eqref{TRS_symm} of the main text. Consequently, for $p=1$ and $p=3$, we have $p+q$ to be odd. This implies that the corresponding current can be finite even when $({\cal T}\chi_{aa;zz}) = \chi_{aa;zz}$, {\it i.e.,} the response can occur in a $\cal T$-symmetric nonmagnetic system. This is consistent since, in conventional metals, the longitudinal magnetoresistance usually follows $B^2$ dependence. From the above equation, we find that $\chi_{aa;zz}$ represents polar, $\cal T$-even response, and the tensor is spatially symmetric in its first and last two indices. Then its Jahn symbol becomes $\rm [V^2] [V^2]$. Using this, we evaluate the symmetry restriction under various symmetry elements, summarized in Table~\ref{table_symmetry2}.

The transverse magnetoconductivity $\propto B^2$ contains terms with $\tau$-scaling of $\tau^0$, $\tau$, $\tau^2$ and $\tau^3$. In general, the response can be represented by 
\be 
j_a = \tau^p \chi_{ab;zz} E_b B_z^q, 
\ee
with $p=\{0,1,2,3\}$ and $q=2$. The $\tau$ and $\tau^3$ dependent terms are $\cal T$-even and spatially symmetric in the current and electric field indices and {\it i.e.,} $\chi_{ab;zz} = \chi_{ba;zz}$. We denote it by $\chi_{[ab];zz}$, with the Jahn symbol being $\rm [V^2] [V^2]$. Conversely, the $\tau$-independent and $\tau^2$ dependent terms are spatially antisymmetric under the exchange of $a\leftrightarrow b$ and $\cal T$-odd (since in this case $p+q$ is even). We denote the response tensor by $\chi_{\{ab\}; zz}$, with the Jahn symbol $ \texttt{a}\rm \{V^2\}[V^2]$. Using these, we analyze the symmetry restriction, which is detailed in Table~\ref{table_symmetry2}.
%
% \begin{table*}[h]
%     \centering
%     \caption{The tick-mark (\cmark) and cross-mark (\xmark) represent that the corresponding elements are allowed and forbidden, respectively.  }
%     \renewcommand{\arraystretch}{1}
%     \setlength\tabcolsep{0.24 cm}
%     \begin{tabular}{c | c c c c }
%        \hline \hline
%          & ${\tau^0}$ & ${\tau^1}$ & ${\tau^2}$  & ${\tau^3}$ 
%          \\ 
%         \hline 
%        $\sigma_{xx}(B^2)$ &  \xmark & \cmark & \xmark & \cmark  \\
%         % $\chi_{xx;z}^{\rm LMC}$ & \xmark & \xmark & \cmark & \xmark & \xmark & \cmark & \cmark & \cmark & \cmark & \xmark & \cmark & \cmark & \xmark & \xmark  \\
%         $\sigma_{yx}(B^2)$  & \cmark & \cmark & \cmark & \cmark    \\
%         \hline \hline
%     \end{tabular}
%     \label{}
% \end{table*}
%
\begin{table*}[h]
    \centering
    \caption{The crystalline point group symmetry restrictions $B$-quadratic longitudinal and transverse conductivities. The tick-mark (\cmark) and cross-mark (\xmark) represent that the corresponding elements are symmetry allowed and forbidden, respectively. }
    \renewcommand{\arraystretch}{1}
    \setlength\tabcolsep{0.24 cm}
    \begin{tabular}{c | c c c c c c c c c c c c c c c c}
       \hline \hline
        Response & ${\cal T}$ & ${\cal M}_x$ & ${\cal M}_y$ & ${\cal M}_z$  & ${\cal C}_{2x}$ & ${\cal C}_{2y}$ & ${\cal C}_{2z}$ &  ${\cal C}_{3z}$ & ${\cal C}_{4z}$  & ${\cal M}_x \cal T$ & ${\cal M}_y \cal T$ & ${\cal M}_z {\cal T}$ & ${\cal C}_{2x}\cal T$ & ${\cal C}_{2y}\cal T$ & ${\cal C}_{2z} \cal T$  %& ${\cal C}_{4z}\cal T$ 
         \\ 
        \hline 
        $\chi_{xx;zz}$ & \cmark & \cmark & \cmark & \cmark & \cmark & \cmark & \cmark & \cmark & \cmark & \cmark & \cmark & \cmark & \cmark & \cmark & \cmark   \\
        
       $\chi_{[yx];zz}$ & \cmark & \xmark & \xmark & \cmark & \xmark & \xmark & \cmark & \xmark & \xmark &\xmark & \xmark & \cmark & \xmark & \xmark & \cmark  \\
       
        $\chi_{\{yx\};zz}$ & \xmark & \xmark & \xmark & \cmark & \xmark & \xmark & \cmark & \cmark & \cmark &\cmark & \cmark & \xmark & \cmark & \cmark & \xmark  \\
        % $\chi_{xx;z}^{\rm LMC}$ & \xmark & \xmark & \cmark & \xmark & \xmark & \cmark & \cmark & \cmark & \cmark & \xmark & \cmark & \cmark & \xmark & \xmark  \\
        
        \hline \hline
    \end{tabular}
    \label{table_symmetry2}
\end{table*}

{\section{Temperature dependence of OMC\%}\label{app_C_new}

% The OMC\% is defined as the ratio $\sigma_{xx}^{\rm OMC}/\sigma_{xx}^{\rm even}$. 
% By symmetry, an odd-in-$B$ longitudinal response is allowed only when time-reversal symmetry (TRS) is broken, and therefore its leading dependence must be proportional to the TRS-breaking parameter in the Hamiltonian. 
% In our case this parameter is the Haldane mass $\Delta$, so $\sigma^{\rm OMC}_{xx}\propto \Delta$ to lowest order. 
% The even-$B$ longitudinal conductivity, by contrast, is a TRS-even quantity dominated by the Drude contribution and the conventional $B^{2}$ magnetoconductivity; these terms remain finite in the TRS-preserving phase and depend only weakly on $\Delta$. 
% Consequently, once the temperature dependence of the order parameter is introduced, the variation of ${\rm OMC}\%$ is controlled almost entirely by the temperature dependence of , and thus directly reflects the behaviour of $\Delta(T)$. 
% Near the critical temperature, where $\Delta(T)$ is small, the odd-parity response collapses as $\Delta(T)\to 0$, whereas the even-$B$ background remains essentially unchanged. 
% Therefore, the OMC\% naturally tracks the temperature evolution of the magnetic order parameter and provides an experimentally accessible probe of intrinsic TRS breaking.

The OMC\% is defined as the magnitude of the ratio $\sigma_{xx}^{\rm OMC}/\sigma_{xx}^{\rm even}$. By symmetry, an odd-in-$B$ longitudinal response is permitted only when time-reversal symmetry (TRS) is broken. Consequently, its leading dependence must be proportional to the TRS-breaking parameter in the Hamiltonian. In contrast, $\sigma_{xx}^{\rm even}$ is a TRS-even quantity dominated by the Drude contribution and the conventional $B^{2}$ magnetoconductivity, both of which remain finite in TRS-preserving materials. Also, these quantities should weakly depend on the TRS-breaking parameter in the Hamiltonian. Consequently, when the temperature dependence of the TRS-breaking parameter is considered, the ${\rm OMC}\%$ inherits its temperature dependence primarily from the $\sigma^{\rm OMC}_{xx}$, which is directly proportional to the TRS-breaking parameter in leading order. Hence, OMC\% closely follows the temperature evolution of the order parameter. This argument is applicable near $T\simeq T_{c}$ where the TRS-breaking parameter is small: the odd-parity response vanishes as $T$ approaches $T_c$, while the even-$B$ background remains minimally affected. As a result, OMC\% provides a direct and experimentally accessible measure of the underlying TRS-breaking amplitude.

% The OMC\% is defined as the ratio of $\sigma_{xx}^{\rm OMC}$ to $\sigma_{xx}^{\rm even}$. From symmetry, the odd-parity magnetoconductivity is allowed only when time-reversal symmetry (TRS) is broken. Therefore, its magnitude must be proportional to the TRS-breaking parameter that enters the Hamiltonian. In our Dirac model, this parameter is the Haldane mass term $\Delta$, which changes sign under ${\cal T}$ and vanishes in the TRS-preserving phase. Hence, $\sigma_{xx}^{\rm OMC}$ necessarily scales as $\Delta$ to leading order. In contrast, the even-$B$ longitudinal conductivity is a TRS-even quantity. It receives contributions such as the Drude term and the conventional $B^{2}$ magnetoconductivity, both of which remain finite even when $\Delta=0$. Thus, the denominator $\sigma_{xx}^{\rm even}$ is essentially sensitive to TRS breaking, while the numerator $\sigma_{xx}^{\rm OMC}$ vanishes continuously as $\Delta \to 0$. As a result, the ratio ${\rm OMC}\%=\frac{\sigma_{xx}^{\rm OMC}}{\sigma_{xx}^{\rm even}}\times100$ inherits the leading TRS-breaking dependence from $\sigma_{xx}^{\rm OMC}$ itself, and therefore tracks the temperature evolution of the magnetic order parameter $\Delta(T)$. This symmetry argument applies around $T\sim T_c$ for small $\Delta(T)$ and explains why the OMC\% closely follows $\Delta(T)$: the numerator is controlled by TRS breaking, while the denominator is dominated by TRS-even classical contributions that vary weakly with $\Delta(T)$.

Having explained why OMC\% should directly depend on the $\Delta(T)$ from the symmetry-based qualitative argument, we now show this explicitly using the valley polarized gapped Dirac model considered in the main text. In the expression of OMC\% of Eq.~\eqref{OMC_per}, the $\sigma^{\rm tot}_{ab}$ is the measured total conductivity and $\sigma^{\rm even}_{ab}$ is the even-$B$ part, including the zero-field Drude contribution. This experimentally accessible ratio removes the dominant Drude and even-$B$ backgrounds, isolating the odd-parity response. To estimate OMC, we use the expression of $\sigma^{\rm even}_{xx} = \sigma^{\rm D}_{xx} + \sigma_{xx}(B^2)$, with $\sigma^{\rm D}_{xx} = \frac{e^2 \tau}{\hbar} \frac{\varepsilon^2 - \Delta^2}{|\varepsilon|}$ being the Drude conductivity. Additionally, we have $\sigma^{\rm OMC}_{xx}= - {\rm sgn}(\varepsilon) \xi \Delta \frac{ \tau e^3  v_F^2}{4\pi \hbar} \frac{2\varepsilon^2-\Delta^2}{\varepsilon^4}B$ [Eq.~\eqref{sigmaxx_B_gr}] and $\sigma_{xx}(B^2) = - \frac{ e^4  v_F^4 }{4\pi } \frac{\tau^3}{\hbar^2}\frac{\varepsilon^2-\Delta^2}{|\varepsilon|^3} B^2 $ [Eq.~\eqref{sigmaB2_xx}]. We have ignored the linear in $\tau$ contributions to $\sigma_{xx}(B^2)$, since this originates from the band geometric quantities, which are very small compared to the conventional classical contribution $\propto \tau^3$. We obtain the OMC to be (valley contributions are summed up) 
\bea
\eta(\varepsilon) = {\rm OMC} = e\hbar v_F^2 B \frac{(2 \varepsilon^2 -\Delta^2)}{(|\varepsilon|^2 - C^2) (\varepsilon^2 - \Delta^2)} \frac{\Delta}{|\varepsilon|} , 
\eea 
where $C= \tau e v_F^2 B \sim \rm 4$~meV, on using $\tau =10^{-12}$~s, $v_F = 2\times10^5$~m/s, and $B=0.1$~T. For low-energy Dirac bands where $\varepsilon$ is typically around tens of meV, $C < \varepsilon$. Furthermore, $\Delta(T)$ is small around the critical temperature, assuming the chemical potential is away from the band edges, we have $\varepsilon \gg \Delta$. Under the experimentally relevant hierarchy $\Delta \ll |\varepsilon| $ and ${C} < \varepsilon$, we obtain the leading order expression of $\eta(\varepsilon)$ to be
\be
\eta(\varepsilon) = e\hbar v_F^2 B \frac{(2\varepsilon^2 -\Delta^2)}{(|\varepsilon|^2 - C^2) (\varepsilon^2 - \Delta^2)} \frac{\Delta}{|\varepsilon|} \approx {2 e \hbar v_F^2 B} \frac{\Delta}{|\varepsilon|^3} + {\cal O}(\Delta^3)~.
\ee
In the Dirac model, the energy-resolved kernel entering the $\eta$ is $ \propto \frac{\Delta}{|\varepsilon|^3}$ to leading order in the TRS-breaking parameter. The finite–temperature $\eta$ is obtained from the energy-resolved kernel $\eta(\varepsilon,\Delta(T))$ by the convolution
\be
\eta(\mu, T)=\int d\varepsilon\;[-\partial_{\varepsilon}f(\varepsilon,\mu,T)]\;\eta(\varepsilon,\Delta(T)),
\ee
where $f^0$ is the Fermi–Dirac distribution function and $\Delta(T)$ is the temperature-dependent order parameter. Using the Sommerfeld regime $k_B T\ll|\mu|$, we have
\be 
\eta(\mu, T) = {2 e \hbar v_F^2 B}  \int d\varepsilon\;(-\partial_{\varepsilon}f^0) \, \frac{\Delta(T)}{\varepsilon^3}
=  {2 e \hbar v_F^2 B} \, \Delta(T) \left[ \frac{1}{\mu^3}+\frac{\pi^2}{6}(k_B T)^2\left.\frac{d^2}{d\varepsilon^2} \left(\frac{1}{\varepsilon^3}\right)\right|_{\varepsilon=\mu}+{\cal O}(T^4) \right].
%=\frac{1}{\mu}+\frac{\pi^2}{3}\frac{(k_B T)^2}{\mu^3}+O(T^4).
\ee
This implies that the dominant contribution to the OMC\% is
\be 
{\rm OMC}\% \approx 100\times\frac{2 e \hbar v_F^2 B}{|\mu|^3}  \Delta(T)~.
\ee
This clearly demonstrates that the dominant temperature dependence of the OMC\% originates from $\Delta(T)$, and thermal smearing only produces small relative corrections of order $(k_BT/\mu)^2$.

Both symmetry considerations and explicit analytical evaluation show that ${\rm OMC}\% \propto \Delta(T)$ for $T\sim T_{c}$. The odd-parity signal, therefore, tracks the temperature evolution of the TRS-breaking order parameter with only weak thermal corrections.  }

\section{Calculation of magnetoconductivity in the presence of Landau levels}\label{app_C}
In this Appendix, we present the details of the calculation for longitudinal collision conductivity in the presence of a quantized Landau level. In the Landau gauge $\bm{A} = (0,xB,0)$, the Hamiltonian~\eqref{Ham_B} can be written as
\begin{align}
    {\cal H}_B^\xi = v_F \left[\xi p_x \sigma_x + (p_y + eBx)\sigma_y \right] + \Delta \sigma_z.
\end{align}
The Hamiltonian is translationally invariant along the \(y\)-direction as \([{\cal H}_B^\xi, p_y] = 0\), which allows the electronic wave function to be written as \(\Psi(x, y) \sim e^{i k_y y} \Phi(x)\). Using this, the eigenvalue problem reduces to ${\cal H}_B^{\xi}\Phi(x)= \varepsilon_n \Phi(x)$ with
\begin{align}\label{ham}
    {\cal H}_B^{\xi}= \frac{v_F \hbar}{l_B} \left[\xi P_x \sigma_x + X \sigma_y \right] + \Delta \sigma_z,
\end{align}
where, the dimensionless $x$-component of momentum 
operator $P_x = -i \partial / \partial(x / l_B)$, position operator 
$X = (x + x_0) / l_B$ with the center of cyclotron orbit at $x = -x_0 = -k_y l_B^2$. Eq.~\eqref{ham} can be written as,
\begin{equation}
    {\cal H}_B^{\xi}=\hbar \omega_c\sqrt{2}
    \begin{pmatrix}
\Delta/\hbar \omega_c\sqrt{2} & -i (a\delta_{\xi,1}+a^\dagger\delta_{\xi,-1}) \\
i (a^\dagger\delta_{\xi,1}+a\delta_{\xi,-1}) & -\Delta/\hbar \omega_c\sqrt{2}
\end{pmatrix},
\end{equation}
where, Kronecker delta function $\delta_{p,q}=0$ for $p\neq q$ and $\delta_{p,q}=1$ for $p=q$. The raising and lowering operators of Harmonic oscillators are defined as $a = ( X + i P_x) / \sqrt{2}$ and $a^\dagger = ( X - i P_x) / \sqrt{2}$, such that $[a, a^\dagger]=1$. Solving the above Hamiltonian, we obtain energy eigenvalues as,
\begin{equation}
\varepsilon_n =
\begin{cases} 
\lambda \sqrt{\Delta^2 + 2n \hbar^2 \omega_c^2}, & n \neq 0, \\ 
-\xi\Delta, & n = 0.
\end{cases}
\end{equation}
% using the ansatz for $n \neq 0$,
% \[
% \Phi_{n,k_y}^{\xi=1}(X) =
% \begin{pmatrix}
%     A \phi_{n-1}(X) \\
%     B \phi_n(X)
% \end{pmatrix},
% \Phi_{n,k_y}^{\xi=-1}(X) =
% \begin{pmatrix}
%     A \phi_{n}(X) \\
%     B \phi_{n-1}(X)
% \end{pmatrix},
% \]
Thus, the wavefunction for $n \neq 0$ can be evaluated as,
\begin{equation}
   \Phi_{n,k_y}^{\xi}(X) = \frac{e^{ik_y y}}{\sqrt{L_y}}
\begin{pmatrix}
    \cos{(\alpha/2)} \left[\phi_{n-1}(X)\delta_{\xi,1}+\phi_{n}(X)\delta_{\xi,-1}\right] \\
    i\sin{(\alpha/2)} \left[\phi_{n}(X)\delta_{\xi,1}+\phi_{n-1}(X)\delta_{\xi,-1}\right]
\end{pmatrix}.
% \Phi_{n,k_y}^{\xi=-1}(X) = \frac{e^{ik_y y}}{\sqrt{L_y}}
% \begin{pmatrix}
%     \cos{(\alpha/2)} \phi_{n}(X) \\
%     i\sin{(\alpha/2)} \phi_{n-1}(X)
% \end{pmatrix}, 
\end{equation}
Here, we have defined $ \phi_n(x) = \frac{1}{\sqrt{2^n n!}}(\frac{1}{\pi l_B^2})^{1/4} H_n\left(x\right)e^{-x^2}$ and $H_n(x)$ is the Hermite polynomial with $\cos{\alpha}=\frac{\Delta}{\varepsilon_n} $. Since the wavefunction is plane wave like in $y$ direction, the normalization is imposed by a finite box of length $L_y$. The wavefunction for $n = 0$ is given by
\begin{equation}
    \Phi_{0,k_y}^{\xi}(X) =\frac{e^{ik_y y}}{\sqrt{L_y}}i \phi_0(X)
\begin{pmatrix}
    \delta_{\xi,1} \\
    \delta_{\xi,-1}
\end{pmatrix}.
\end{equation}
% \textcolor{red}{What is $L_y$? Specify every parameter defined.... }

The longitudinal conductivity arises mainly due to the scattering of cyclotron orbits from the charge impurities. This contribution is also known as collisional conductivity. In the presence of perpendicular electric and magnetic fields, the collisional conductivity in the low temperature regime can be evaluated by using the following formula~\cite{Vasil_jmp82, Vasil_prb92}
\begin{equation}
    \sigma^{\text{col}}_{xx} = \frac{\beta_T e^2}{2 S_0} \sum_{\zeta, \zeta'} f_\zeta \left(1 - f_{\zeta'}\right) W_{\zeta \zeta'} \left(x_\zeta - x_{\zeta'}\right)^2\label{ab},
\end{equation}
here, $\zeta \equiv \{\xi, \lambda, n, k_y \}$ denotes all the quantum indices. The average value of the $x$-component of the position operator in state $|\zeta\rangle \equiv \Phi_{n,k_y}^{\xi}(X)$ is given by $x_\zeta = \langle \zeta | x | \zeta \rangle$, which can be evaluated as $x_\zeta = \langle \zeta |-x_0 | \zeta \rangle + \langle \zeta | l_B X | \zeta \rangle = -k_y l_B^2 $ thus $(x_\zeta - x_{\zeta'})^2 = (q_y l_B^2)^2$ with $k'_y - k_y = q_y$. The scattering rate between states $\zeta$ and $\zeta'$ is given by,
\begin{equation}\label{scat rate}
    W_{\zeta \zeta'} = \frac{2\pi n_{\rm im}}{S_0\hbar} \sum_q |U_q|^2 |F_{\zeta \zeta'}(\eta)|^2 \delta(\varepsilon_\zeta - \varepsilon_{\zeta'}) \delta_{k_y, k'_y + q_y}.
\end{equation}
The form factor is defined as,
\be
    F_{\zeta \zeta'} = \langle \zeta | e^{i \mathbf{q} \cdot \mathbf{r}} | \zeta'\rangle =\langle\Phi_{n, k_y}^{\xi} | e^{i \mathbf{q} \cdot \mathbf{r}} | \Phi_{n', k_y'}^{\xi} \rangle.
\ee
% \textcolor{red}{As for the evaluation of $\sigma^{\text{col}}_{xx}$, now there are no restrictions on the intralevel scattering, i.e., $n' = n$ is allowed, while $n' \neq n$ is forbidden for elastic scattering, since the eigenvalues do not depend on $k_y$.} \textcolor{cyan}{ This feels disconnected and not clear---add some more details or make it clear....} 
Now to explicitly evaluate $F_{\zeta\zeta'}$, we restrict ourselves to consider only the intra-band ($\lambda'= \lambda$) and intra-level ($n' = n$) scattering because of the presence of the term $\delta(\varepsilon_\zeta - \varepsilon_{\zeta'}) $ in $W_{\zeta \zeta'}$.
% \begin{widetext}
Using Taylor expansion of Hermite polynomial $H_n(x+y) = \sum_{k=0}^{n} \binom{n}{k} H_k(x) (2y)^{n-k}$ and orthogonality property $\int_{-\infty}^{\infty} e^{-x^2} H_m(x) H_n(x) \, dx = \sqrt{\pi} \, 2^n n! \, \delta_{mn} $ we can get,
\begin{equation}
\langle \phi_{n}(X) | e^{i q_x x} | \phi_{n}(X') \rangle=e^{-i\Theta - \frac{\eta}{2}}\frac{1}{2^n n! \sqrt{\pi}} \int e^{-x^2} H_n\left( x + i \nu \right) H_n\left( x + i  \nu^* \right) dx= e^{-i\Theta - \frac{\eta}{2}}L_n(\eta),
\end{equation}
where $\nu = l_B(q_x + i q_y)/2$, $\Theta = l_B^2 q_x (k_y + q_y/2)$, and $L_n(\eta)$ is the Laguerre polynomial. Using the above equation, the form factor is obtained to be
\begin{equation}
F_{\zeta \zeta'} =
\frac{1}{2}\,\delta_{k_y',\,k_y-q_y}\, {e}^{-i\Theta-\eta/2}
\begin{cases}
\left(1+\dfrac{\Delta}{\varepsilon_n}\right)L_{n-1}(\eta)
+\left(1-\dfrac{\Delta}{\varepsilon_n}\right)L_{n}(\eta) & \text{for} ~~\xi=+1,\\
 \left(1+\dfrac{\Delta}{\varepsilon_n}\right)L_{n}(\eta)
+\left(1-\dfrac{\Delta}{\varepsilon_n}\right)L_{n-1}(\eta) & \text{for} ~~\xi=-1 .
\end{cases}
\end{equation}
Substituting $\sum_q \to \frac{\Omega}{(2\pi)^2} \int q \, dq \, d\phi$ and form factor expression in Eq. \eqref{scat rate} one can obtain scattering rate as,
\begin{equation}
    W_{\zeta \zeta'} = \frac{n_{\rm im} U_0^2}{8\pi \hbar k_s^2}
    \begin{cases}
        \displaystyle \int e^{-\eta}
        \left[\left(1+\frac{\Delta}{\varepsilon_n}\right)L_{n-1}(\eta)
        +\left(1-\frac{\Delta}{\varepsilon_n}\right)L_{n}(\eta)\right]^2
        \delta(\varepsilon_\zeta - \varepsilon_{\zeta'}) \, q\, dq\, d\phi, & \text{for}~\xi=+1,\\[1.2em]
        \displaystyle \int e^{-\eta}
        \left[\left(1+\frac{\Delta}{\varepsilon_n}\right)L_{n}(\eta)
        +\left(1-\frac{\Delta}{\varepsilon_n}\right)L_{n-1}(\eta)\right]^2
        \delta(\varepsilon_\zeta - \varepsilon_{\zeta'}) \, q\, dq\, d\phi, & \text{for}~\xi=-1.
    \end{cases}
\end{equation}
Assuming Lorentzian broadening of the $\delta$-function, we can write 
$ \delta(\varepsilon_\zeta -\varepsilon_{\zeta'}) = \Gamma_0/\pi[(\varepsilon_\zeta -\varepsilon_{\zeta'})^2 + \Gamma_0^2], $ 
where $\Gamma_0$ is the broadening parameter. It may depend on the magnetic field, the quality of the samples, etc. For intra-Landau level and intraband scattering, we may further write $\delta(\varepsilon_\zeta -\varepsilon_{\zeta'})=1/\pi\Gamma_0. $ By replacing $\sum_{\zeta,\zeta'}\to\frac{S_0}{(2\pi l_B^2)} \sum_{n}$, and $(x_\zeta - x_{\zeta'})^2 = q_y^2 l_B^4 = (q \sin \phi)^2 l_B^4$ in Eq.~\eqref{ab} one can obtain longitudinal magnetoconductivity as,
\begin{equation}\label{conductivity}
    \sigma_{xx} = \frac{e^2}{h} 
\frac{n_{\rm im} U_0^2}{2\pi k_s^2 l_B^2 \Gamma_0}  \beta_T
\sum_{n, \lambda} I_n^{\xi} f_{\zeta} (1 - f_{\zeta'}),
\end{equation}
% \textcolor{cyan}{ Here, I took $f_{n, \lambda} $ while earlier it was $f_{\zeta} $. I have to show energy is the same though there are two states $\zeta,\zeta'$ through $k_y$. Suggest me correct way to write it.} \textcolor{red}{write it down first with clarity and deatils then we will review}
with
\begin{equation}
    I_n^{\xi} =
    \begin{cases}
        \displaystyle \int \eta e^{-\eta}
        \left[\left(1+\frac{\Delta}{\varepsilon_n}\right)L_{n-1}(\eta)
        +\left(1-\frac{\Delta}{\varepsilon_n}\right)L_{n}(\eta)\right]^2
        d\eta, & \text{for}~\xi=+1,\\[1.2em]
        \displaystyle \int \eta e^{-\eta}
        \left[\left(1+\frac{\Delta}{\varepsilon_n}\right)L_{n}(\eta)
        +\left(1-\frac{\Delta}{\varepsilon_n}\right)L_{n-1}(\eta)\right]^2
        d\eta, & \text{for}~\xi=-1.
    \end{cases}
\end{equation}
Using the orthogonality of the Laguerre polynomials  
$ \int_{0}^{\infty} e^{-x} L_m(x) L_n(x) \, dx = \delta_{mn} $ and the recurrence relation  
$ x L_n(x) = (2n+1) L_n(x) - (n+1) L_{n+1}(x) - n L_{n-1}(x)$, we obtain
\begin{equation}
    I_n^{\xi}=2n - \xi\frac{4\Delta}{\varepsilon_n} + \frac{6n \Delta^2}{\varepsilon_n^2}.
\end{equation}
Since we used $n=n',~\lambda=\lambda'$, and the LL energy depends on only these two quantum numbers, in the low temperature limit, the product $ \beta_T f_{\zeta} (1 - f_{\zeta'})$ can be replaced by $\delta(\varepsilon_n - \mu)$. Now, Eq. \eqref{conductivity} can be written as,
\begin{equation}\label{final}
    \sigma_{xx}=\frac{e^2}{h} 
\frac{n_{\rm im} U_0^2}{2\pi k_s^2 l_B^2 \Gamma_0}
\sum_{n, \xi} \left[2n - \xi\frac{4\Delta}{\varepsilon_n} + \frac{6n \Delta^2}{\varepsilon_n^2}\right]\delta(\varepsilon_n - \mu).
\end{equation}
We mention that the above conductivity expression is consistent with the Ref.~\cite{Vasi_prb12}. Similarly, one can obtain the magnetoconductivity contribution from the zeroth Landau level to be given by Eq.~\eqref{sigma_ultra}.
% \end{widetext}
\twocolumngrid
\bibliography{refs}

\end{document}